\documentclass[12pt]{article}
\usepackage{epsfig}
\usepackage{amsmath,amssymb,amsthm,amscd}

\def\IP{{\mathbb P}}

\def\ZZ{{\mathbb Z}}

   \let\S=\Sigma
\def\2{{1\over2}}
\let\<=\langle \let\>=\rangle
\def\new#1\endnew{{\bf #1}}
\def\ifundefined#1{\expandafter\ifx\csname#1\endcsname\relax}
\ifundefined{draftmode}\else    \input draftmode        \fi
\let\Msize=\footnotesize             
\def\BM{\Msize\begin{matrix}}           \def\EM{\end{matrix}}
\def\MN M:#1 #2 N:#3 #4 {{(#1_{#2},#3_{#4})}}
\def\MNH M:#1 #2 N:#3 #4 H:#5,#6 [#7]{{(#1_{#2},#3_{#4})^{#5,#6}_{#7}}}

\newcommand{\ds}{\displaystyle}

\def\dd{\mathrm{d}}

\newcommand{\tr}{{\rm Tr}}

\newcommand{\be}{\begin{equation}}
\newcommand{\ee}{\end{equation}}
\newcommand{\bea}{\begin{eqnarray}}
\newcommand{\eea}{\end{eqnarray}}

\def\tr{\hbox{tr}}

\begin{document}
\begin{titlepage}
{}~
\hfill\vbox{
\hbox{}
}\break

\rightline{hep-th/0605195}
\rightline{MAD-TH-06-5} \vskip 1cm

\centerline{\Large \bf
Holomorphic Anomaly in Gauge Theories and Matrix Models}  \vskip 0.5 cm
\renewcommand{\thefootnote}{\fnsymbol{footnote}}
\vskip 30pt \centerline{ {\large \rm Min-xin Huang
\footnote{minxin@physics.wisc.edu} and Albrecht Klemm
\footnote{aklemm@physics.wisc.edu} }} \vskip .5cm \vskip 30pt
\centerline{\it Department of Physics, University of Wisconsin}
\centerline{\it Madison, WI 53706, U.S.A.}

\setcounter{footnote}{0}
\renewcommand{\thefootnote}{\arabic{footnote}}
\vskip 100pt
\begin{abstract}
We use the holomorphic anomaly equation to solve the gravitational corrections to
Seiberg-Witten theory and a two-cut matrix model, which is related by the Dijkgraaf-Vafa
conjecture to the topological $B$-model on a local Calabi-Yau manifold. In both cases we
construct propagators that give a recursive solution in the genus modulo a
holomorphic ambiguity. In the case of Seiberg-Witten theory the gravitational
corrections can be expressed in closed form as quasimodular functions of $\Gamma(2)$.
In the matrix  model we fix the holomorphic ambiguity up to genus two. The latter
result establishes the Dijkgraaf-Vafa conjecture at that genus and yields a new
method for solving the matrix model at fixed genus in closed form
in terms of generalized hypergeometric functions.

\end{abstract}

\end{titlepage}
\vfill
\eject

%%%%%%%%%%%%%%%%%%%%%%%%%%%%%%%%%%%%%%%%%%%%%%%%%%%%%%%%%%%%%

\newpage

\baselineskip=16pt

\tableofcontents

%\newpage

\section{Introduction}

The holomorphic anomaly equations, discovered in\cite{BCOV} in a
world-sheet analysis of a topological twisted $\sigma$-model known
as B-model, are a generalisation of Quillens anomaly to higher
genus and more general world-sheet theories. Gauge theory are
embedded in string theory and in the $N=2$ context the latter can
be obtained in double scaling decoupling limit. The holomorphic
anomaly equations commute with that decoupling limit and the
recursive procedure which determine the higher
${F}^{(g)}(\tau,\bar \tau)$ \cite{BCOV}, which describe certain
terms in the coupling to gravity,  can be build up entirely from
gauge theory quantities. In section \ref{SWsection} the
corresponding topological partition of the 4d $N=2$ SUSY gauge is
determined by solving the holomorphic anomaly recursively, up to
finitely many terms, which have to be fixed by analyzing the
boundary behaviour of the ${F}^{(g)}(\tau,\bar \tau)$. The
properties of $F^{(g)}(\tau, \bar \tau)$ are to a large extend
fixed by the modular group of the corresponding Seiberg-Witten
curve and we are able to write global expressions them in terms of
``almost holomorphic'' modular functions of this group. Various
holomorphic limits are readily taken from our expressions and
provide conjectural solutions to the unitary matrix model
\cite{DV2}.

Riemann surfaces $\Sigma_g$ of all genus can be embedded in
noncompact Calabi-Yau threefolds $X$, i.e. the local limit of the
$B$-model exist for an arbitrary Riemann surface. There are
choices in this embedding, which affect the reduction of the
holomorphic $(3,0)$ form from $X$ to one form $\lambda$ on
$\Sigma_g$, which is a key datum of the local $B$-model limit.
Except for the product case, $\lambda$ will be a meromorphic form
with residua and eventually boundary data on open $\Sigma_{g,h}$.
In  the pure SW-case, there are  poles with vanishing residua. In
the massive SW-case and other local geometries, one one would have
parameters associated with the non-vanishing residua. In section
\ref{DVsection1} we apply the holomorphic anomaly equations to
$N=1$ 4d gauge theory in a geometry proposed by \cite{CIV}. This
is a case where $\lambda$ has essential singularities or
differently put one deals with open Riemann surfaces. According to
the Dijkgraaf-Vafa correspondence~\cite{DV1} the holomorphic
anomaly equations provide here in a particular region solutions to
the large $N$ expansion of a complex matrix model, which is the
solution to an open string problem. This has a description in the
$A$ and the $B$ language as summarized following commuting diagram
of duality relations
\begin{center}
\begin{figure}[htbp]
~~~~~~~~~~~\epsfig{file=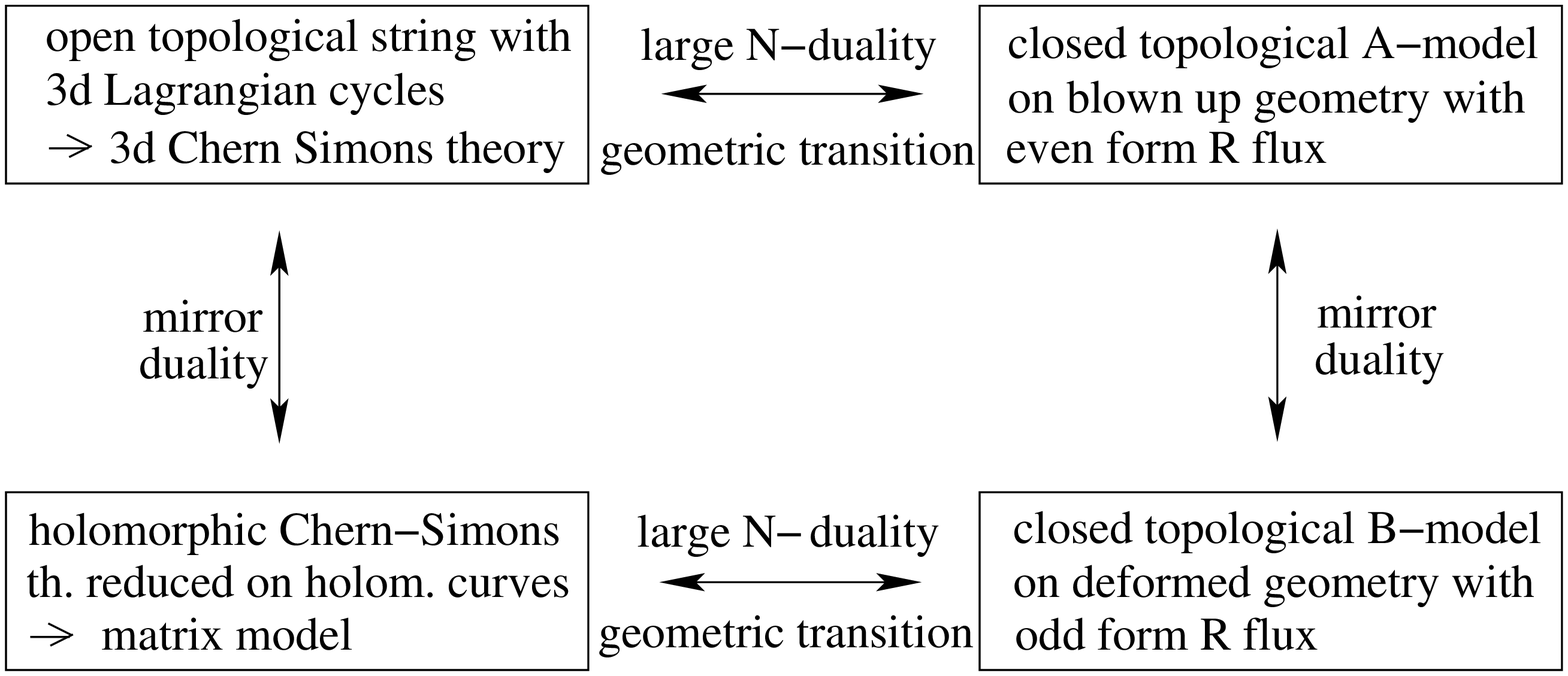,width=.8\textwidth}
\label{fig:1}
\end{figure}
\end{center}
The Dijkgraaf-Vafa conjecture, relevant to calculate effective
terms in $\mathcal{N}=1$ supersymmetric theories in 4d, involves
non-compact Calabi-Yau geometries, which are in particular not
toric. Explicit test of the conjecture are therefore difficult and
have been done only at tree level~\cite{DV1} and at genus
one~\cite{KMT, DST}. In the genus one case it is found that the
free energy for multi-cut matrix models can be written in closed
form \cite{KMT, Vasiliev:2005qj}. We set the B-model formalism up
to get recursively closed expressions at all genus and check the
Dijkgraaf-Vafa correspondence of the first non-trivial example,
the two cut case, explicitly at genus two.

As in the $N=2$ case, we make a detailed analysis of the moduli
space and the modular transformation of the periods in Appendix
\ref{monodromy}, which enables us to solve the theory at various
regions in the moduli space. Different from in the $N=2$ case
however there is a divisor with essential singularities of the
periods in the moduli space, where the perturbative description
breaks down. From the topological string point of view we obtain
solutions for a new class of non-toric, non-compact Calabi-Yau
geometries, which have no point of maximal unipotent monodromy.

\section{Gravitational corrections to N=2 Seiberg-Witten Gauge
theory} \label{SWsection}

In this section we introduce a B-model, which calculates the
higher genus space-time instantons of $N=2$ Seiberg-Witten gauge
theory. The genus $g$ generating function ${\cal
F}^{(g)}(a)$ of these space-time instantons describe
gravitational couplings to the gauge theory~\cite{Moore:1997pc}.
The coefficients of
\begin{equation}
F=\sum_{g=0}^\infty \lambda^{2g-2}{\cal F}^{(g)}(a)
\end{equation}
can be calculated iteratively in the instanton number, $1/a^4$
powers below, using localisation in the moduli space of gauge theory
\cite{Nekrasov:2002qd,Flume:2002az,Nekrasov:2003rj,Nakajima2003} or worldsheet
instantons \cite{KMT,Aganagic:2003db}. Global properties  under
the monodromy group, which should be central in the solution of the
theory, remain obscure in these approaches. We find that studing 
the global properties enables us to solve the B-model completly and in particular fix
the holomorphic ambiguity. This yields closed modular expressions which 
determine the instanton contributions to all orders in the instanton number, 
but  are iteratively in the genus i.e. in $\lambda$. We finish the section 
with some speculations how to obtain closed expressions in $\lambda$ as well.

\subsection{Modular properties of the genus zero and genus one sector}

We focus on simplest case of $SU(2)$ gauge theory without matter.
Generalizations to other gauge groups and matter spectra with
asymptotic freedom are certainly possible. The monodromy group of pure
$SU(2)$ Seiberg-Witten theory is
$\Gamma(2)\in \Gamma_0={\rm SL}(2,\mathbb{Z})$ generated by \cite{Seiberg:1994rs}
\begin{equation}
M_\infty=PT^{-2}, \qquad M_1=S T^2 S \ ,
\label{Gamma2}
\end{equation}
where $S=\left(\begin{array}{rr} 0& -1\\ 1& 0\end{array}\right)$
and $T=\left(\begin{array}{rr} 1& 1\\ 0& 1\end{array}\right)$ are
the generators of  $\Gamma_0$ and
$P=\left(\begin{array}{rr} -1& 0\\ 0& -1\end{array}\right)$.
Using  modular properties as well as the B-model holomorphic
anomaly we are able to express the ${\cal F}^{(g)}$ in terms of $\Gamma(2)$ forms 
and the quasi-modular form of degree two $E_2$.

The natural embedding of gauge theory in type IIB string theory
\cite{KKV} is the explanation that the higher genus worldsheet
technique of \cite{BCOV} applies to the space-time instanton
calculation. More precisely as found in \cite{KKV} IIB theory on
the local Calabi-Yau space ${\cal O}(-2,-2)\rightarrow \IP^1\times
\IP^1$ has a double scaling limit in the two complexified K\"ahler
parameters $t_1,t_2$ of $\IP^1\times \IP^1$ in which ${\cal
F}^{(g)}(t_1,t_2)$ approaches ${\cal F}^{(g)}(a)$. What we find
below is that the holomorphic anomaly equation make sense in the
limit and can be directly viewed as property of the gauge theory.

The Picard-Fuchs equation for the periods $a=\int_a \lambda$ and $a_D=\int_b \lambda$
of the elliptic curve \cite{Seiberg:1994rs}
\begin{equation}
y^2=(x-u)(x-\Lambda^2)(x+\Lambda^2)
\label{curve1}
\end{equation}
with meromorphic differential $\lambda={\sqrt{2}\over 2 \pi }{ y\over x^2-1}\dd x$
\begin{eqnarray}
\label{Picard-Fuchs}
\frac{\partial^2 a}{\partial^2 u}=\frac{a}{4(1-u^2)}\
\end{eqnarray}
allows to calculate the genus zero prepotential using the relation
${\partial {\cal F}^{(0)}\over  \partial a}= a_D$ up an irrelevant
constant. Here for convenience we have set the Seiberg-Witten
scale $\Lambda=1$, which can be easily recovered by dimensional
analysis. The details of the calculation that leads to explicit
expressions for the periods can be found\footnote{In
\cite{Klemm:1995wp} the isogenous curve $y^2=(x^2-\tilde
u)^2-\Lambda^4$ with the meromorphic differential $\lambda={i
\sqrt{2}\over 4 \pi} 2 x^2 {\dd x\over y}$ is used. This means
that $a^{SW}=a^{KLT}$, $a_D^{SW}=a_D^{KLT}/2$,
$\tau_D^{SW}=\tau_D^{KLT}/2$ and  the monodromy is $\Gamma_0(4)$
instead of $\Gamma(2)$.} in \cite{Klemm:1995wp}. The elliptic
curve (\ref{curve1}) with $\Gamma(2)$ monodromy has the
$j$-function $j(\tau)={(3 + u^2)^3\over 27 (u^2-1)^2}$, which
allows to write~\cite{Nahm:1996di}
\begin{equation}
u(\tau)={c+d\over b}\ ,
\label{ut}
\end{equation}
as the Hauptmodul of $\Gamma(2)$ in terms of the ratio $\tau=\int_b {\dd
x\over y}/\int_a {\dd x\over y}=-\frac{1}{4\pi i}{\partial^2 {\cal F}^{(0)}
\over\partial^2 a}$ of the periods over the holomorphic one-form. The
latter are solution of the $_2F_1$ hypergeometric differential equation
$\frac{\partial^2 a}{\partial^2 u}=\frac{2 u }{(1-u^2)}\frac{\partial a}
{\partial  u}+\frac{a}{4(1-u^2)}$, which implies that the inverse relation to
(\ref{ut}) can be written in terms of a Schwarz-triangle function (see e.g. \cite{Klein})
\begin{equation}
\tau(u^2)=s(\frac{1}{4}, \frac{1}{4},1;u^2):={i (1-u^2)^{\frac{1}{4}}
\over 2 (1-i)} {_2F_1(\frac{1}{4}, \frac{1}{4},1;1-u^2)\over
_2F_1(\frac{1}{4}, \frac{1}{4},1;\frac{1}{1-u^2})}\ .
\end{equation}
We defined in (\ref{ut}) $b=\theta_2^4$, $c=\theta_3^4$,
$d=\theta_4^4$ with the relation $b+d=c$. Using the modular transfomation 
\begin{equation}
T: \begin{array}{rl} b&\rightarrow -b\\
                     c&\leftrightarrow d

   \end{array}
\qquad
S: \begin{array}{rl} b&\rightarrow -\tau^2 d\\
                     d&\rightarrow -\tau^2 b\\
                     c&\rightarrow -\tau^2 c

\end{array}
\label{modtrans}
\end{equation}
and (\ref{Gamma2}) it is immediate that $u(\tau)$ is invariant
under $\Gamma(2)$.  For further reference let us note that the
discriminant of the elliptic curve is given by $u=\infty$ and
\begin{equation}
\Delta=u^2-1=4 {cd\over b^2}=0\ . \label{disc}
\end{equation}

The K\"ahler potential in rigid  special geometry is given by (see
\cite{Ferrara:2006js} for a recent review)
\begin{equation}
K=i (X^I \bar F_{\bar I}- {\bar X}^{\bar I} F_I) \ .
\end{equation}
For the Seiberg-Witten curve $X^1=a$ and $\tau :={\partial^2 \over \partial^2 a} F$
so that the metric becomes
\begin{equation}
g_{a\bar a}=\partial_a \bar \partial_{\bar a} K=2 \tau_2\ .
\end{equation}
The genus one amplitude is obtained by integrating the genus holomorphic
anomaly  using  the special geometry relation~\cite{Bershadsky:1993ta}
\begin{equation}
\label{anholF1}
F^{(1)}=-\log(\sqrt{\tau_2} \eta \bar \eta)\ .
\end{equation}
Note that the holomorphic ambiguity is fixed by requiring that $F^{(1)}$ 
is invariant under ${\rm SL}(2,\ZZ)$ transformations and regular 
inside the fundamental domain  $\cal H$. Indeed $\eta$ is the unique modular 
form of weight ${1\over 2}$ so that $F^{(1)}$ regular inside $\cal H$.

In the holomorphic limit\footnote{Because of the volume of the
diffeomorphism group generated by the Killing vector field on $T^2$ we need
regularization of an infinite constant in that limit. Only the
immediately well-defined $\partial_{\tau} F^{(1)}$ plays a r\^ole
in the following. To define that limit consider $\tau$ and $\bar \tau$ 
as independent variables.} $\bar \tau\rightarrow \infty$, we
get~\cite{Bershadsky:1993ta}
\begin{equation}
\mathcal{F}^{(1)}=-\log(\eta(\tau))=-\frac{1}{2}\log(\frac{\partial
a}{\partial u})-\frac{1}{12}\log(u^2-1)\ .
\label{F1}
\end{equation}
This  agrees with \cite{Moore:1997pc}. The form of
$\mathcal{F}^{(1)}$ in terms of $u$ and $\frac{\partial
a}{\partial u}$ follows in the rigid limit from
\cite{Bershadsky:1993ta} and was observed in \cite{KMT}. Using
(\ref{F1}) and $\eta^{12}=2^{-4} bcd$ one gets
\begin{equation}
{\partial a \over \partial u}= {\theta_2^2\over 4}\ .
\label{dadu}
\end{equation}
Further  with (\ref{ut}) and  the formulas for the derivatives of
the $\theta$ functions we find
\begin{equation}
\label{Caaa}
{\cal F}^{(0)}_{,3}(\tau):=
{\partial^3 {\cal F}^{(0)}\over \partial^3 a} =
\frac{\partial \tau}{\partial a}= {8 \sqrt{b} \over c d}={1\over \Delta} \left(\partial u\over \partial a\right)^3 \ .
\end{equation}
This can alternatively derived by taking twice the derivative w.r.t. 
to $a$ in the Matone relation $\mathcal{F}^{(0)}-\frac{1}{2}a\frac{\partial
\mathcal{F}^{(0)}}{\partial a}+u=0$~\cite{Matone}.
Using the Picard-Fuchs (\ref{Picard-Fuchs}) equation and (\ref{dadu}) we
can write the period as\footnote{To keep the notation
simpler we supress normalisation factors of ${1\over 2 \pi i}$ from
the periods, which make them an integral representation of (\ref{Gamma2}).}
\begin{equation}
a(\tau)={1\over 3 \theta_2^2}(E_2+ c + d)\ .
\label{a-period}
\end{equation}
Closely related expressions for the SW-periods of elliptic curves
appear in~\cite{Moore:1997pc} and for  more  general gauge groups
in~\cite{marinoperiods}.

The strong coupling duality element $S:\tau\rightarrow -{1\over \tau}$ of $N=4$
Super-Yang-Mills theory is not trivially realized in pure $N=2$ Seiberg-Witten theory.
It relates the gauge instanton expansion in the region of asymptotic freedom
${1\over u},{1\over a}\rightarrow 0$  to the magnetic $U(1)$ which is weakly
coupled at the  magnetic monopole point $z=(u-1)/2,a_D\rightarrow 0$. The gauge
coupling of the asymptotic free theory is determined by $\tau(a)$, while the
one of the magnetic $U(1)$ is determined by $\tau_D(a_D)$ with the relation
\begin{equation}
\label{s-duality}
\tau_D=-{1\over \tau}
\end{equation}

We note that the one-loop amplitude (\ref{anholF1}) is $S$-duality invariant. The
holomorphic limit at the monopole points is taken by $\bar \tau_{D}\rightarrow \infty$.
Therefore one has
\begin{equation}
\begin{array}{rl} 
\mathcal{F}^{D (1)}(a_D)=&\ds{-\log(\eta(\tau_D))=-\frac{1}{2}\log(\frac{\partial a_D}
{\partial u})-\frac{1}{12}\log(u^2-1)}
\\[3 mm]
=& \ds{\frac{\log (2)}{3} -\frac{i }{6}\, \pi  - \frac{\log (\tilde a_D)}{12} - 
\frac{\tilde a_D}{2^5} - \frac{3\, {\tilde a_D}^2}{2^9} - 
\frac{19\, {\tilde a_D}^3}{2^{12} 3}+ {O(\tilde a_D)}^4}
\end{array}
\label{FD1}
\end{equation}
We derive, similar as (\ref{a-period}) was obtained, from the first line 
of  (\ref{FD1}) a formula for the second period
\begin{equation}
a_D(\tau_D)=-{i\over 3 \theta_4^2}(E_2- b- c)\ .
\label{ad-period}
\end{equation}
This can then be inverted to obtain the series expansion in the second 
line of (\ref{FD1}),  where we rescale 
\begin{equation}
\tilde a_D:=i {a_D\over 2}\ .
\end{equation}
It is naturally to define an anholomorphic period
\begin{equation}
A(\tau, \bar \tau) ={1\over 3 \theta_2^2}(\hat E_2(\tau,\bar
\tau)+c+d),
\label{anholperiod}
\end{equation}
with the anholomorphic weight one form 
\begin{equation}
\hat E_2(\tau,\bar \tau):= E_2(\tau)+ {6 i \over \pi(\bar \tau-\tau)}\ .
\label{hatE2}
\end{equation}
Different then the Eisenstein series $E_4$ and $E_6$, holomorphic modular forms
of weight $4$ and $6$ which generate the ring of holomorphic forms of $\Gamma_0$,
the holomorphic $E_2$ transforms not quite as modular form of weight $1$,
but rather quasi-modular under $\Gamma_0$, i.e. with a shift
\begin{equation}
E_2\left(a\tau + b\over c \tau + d\right)=(c\tau + d)^2 E_2(\tau) +{12\over 2\pi i}
c ( c\tau+ d)\ .
\label{shift}
\end{equation}
However the anholomorphic piece in  (\ref{anholperiod}) cancels the shift so that 
$\hat E_2(\tau,\bar \tau)$ transforms as a form of weight $1$.

It follows from (\ref{anholperiod}) that $a(\tau)=\lim_{\bar \tau\rightarrow \infty} A(\tau,\bar \tau)$ and
$a_D(\tau_D)=\lim_{\bar \tau_{D}\rightarrow \infty} \frac{1}{\tau_D} A(-{1\over \tau_D},-{1\over \bar \tau_D})$.
Morover S-duality transformations (\ref{modtrans}) allow us to write $a$ and $a_D$ as functions
of $\tau$ or $\tau_D$. In particular we can integrate (\ref{a-period}) w.r.t. $a_D$ in 
the magnetic phase 
\begin{equation}
{\cal F}^{D (0)}=\frac{1}{2}{\tilde a_D}^2 \log\left( -{\tilde a_D\over 2}\right)+ 
4 \tilde a_D - \frac{{3\tilde a_D}^2}{2^2}+\frac{{\tilde a_D}^3}{2^4} + 
\frac{5\, {\tilde a_D}^4}{2^9} + \frac{11\, {\tilde a_D}^5}{2^{12}} + O(\tilde a_D^6)\ .
\end{equation}

\subsection{Propagators and integrating the holomorphic anomaly equations}

From the genus zero and genus one amplitude one can derive the propagator \cite{BCOV}.
With the simplification for local $B$-model explained in \cite{Klemm:1999gm} we obtain
\begin{equation}
\label{anholSaa} S^{aa}=S={2 {F}^{(1)}_{,1} \over { F}^{(0)}_{,3}}=
{1 \over 24} \left(E_2(\tau)- {3\over \pi
\tau_2}\right)=:{1 \over 24}\hat E_2(\tau,\bar \tau) \ ,
\end{equation}
which in the holomorphic limit becomes
\begin{equation}
\label{holSaa} {\cal S}^{aa}={\cal S}={2 {\cal F}^{(1)}_{,1} \over
{\cal F}^{(0)}_{,3}}={1 \over 24} E_2(\tau) \ .
\end{equation}
This quantity contains the contribution from the boundaries of the 
WS moduli space in the topological string theory. It is 
closely related to $T_{k,l}$ the quantity that arises if one considers 
correlators of integrated two-form operators constructed
from the descent equation in the gauge theory on four manifolds.
More precisely intersections of the  correponding  2-cycles require contact terms
which are  $T_{k,l}= S^{a_i a_j} {\partial u_k\over \partial a_i}
{\partial u_l\over \partial a_j}$, see \cite{marcosreview}.

Equations (\ref{Caaa},\ref{anholF1},\ref{anholSaa}) define the data needed to recursively
solve the $B$-model \cite{BCOV}. Using the fact that the formalism of integrating
the holomorphic anomaly equations commutes with the double scaling
limit taken to obatin the gauge theory \cite{KKV} and power counting
in the propagator  in the Feynmann rules of \cite{BCOV} we obtain
the following general result
\begin{equation}
F^{(g)}(\tau, \bar \tau)=
X^{g-1} \sum_{k=0}^{3(g-1)} \hat E_2^{k}(\tau,\bar \tau) c^{(g)}_k(\tau)\ .
\label{swsu2ambiguity}
\end{equation}
Here we defined $X= {b\over 1728 c^2 d^2}={1\over {108 (u^2-1)^2 b^3}}$,
which transforms as a weight $-3$ object 
$X({a \tau + b\over c \tau +d})\rightarrow {1\over (c \tau +d)^6} X(\tau)$
under $\Gamma(2)$. $F^{(g)}$ is invariant under $\Gamma(2)$, which
implies that $c^{(g)}_k$ are homogeneous of weight $3(g-1)-k$ in $(b,c,d)$. Further
conditions on $c^{(g)}_k$ come from regularity at $u=0$ and a gap 
condition at the conifold $u=1$ as dicussed below.

We obtain the holomorphic limits of the expansion in the asymptotic freedom  and the strong
coupling region  as
\begin{equation}
{\cal F}^{(g)}(a)=\lim_{\bar \tau \rightarrow\infty}  F^{(g)}(\tau, \bar \tau),
\quad {\rm and} \quad  {\cal F}^{D(g)}(a_D)=\lim_{\bar \tau_{D}\rightarrow\infty}
F^{(g)}(-{1\over \tau_D},-{1\over \bar \tau_D})\ 
\label{limits}
\end{equation}
and the $a$ or $a_D$ ($\tilde a_D$) expansion are obtained by inverting 
(\ref{a-period}) or (\ref{ad-period}).
We note that the leading behaviour of electric and magnetic expansion 
in these parameters is
\begin{equation}
{\cal F}^{ (g)}(a)\sim {(-1)^g  B_{2g } \over g (2 g-2) (2 a)^{2g -2}}\quad {\rm and}\quad
{\cal F}^{D(g)}(a_D)\sim {B_{2g } \over 2 g (2 g-2){\tilde a_D}^{2g -2}}\
\label{asym}
\end{equation}
respectively. The first asymptotic behaviour can be derived in
the gauge theory limit of type II theory on ${\cal O}(-2,-2)\rightarrow
\IP^1\times \IP^1$. More precisley one uses in
the Gopakumar-Vafa expansion $\frac{1}{\left(2 \sin {m\lambda \over 2}\right)^2}
=\sum_{g=0} \lambda^{2g-2}(-1)^{(g-1)} \frac{B_{2g}}{2 g (2g-2)!} m^{2g-2}$ and
the multiplicity $n^{(g)}_{m,0}=\delta_{g,0} \delta_{m,0}=-2$ of BPS states
corresponding to constant maps on one $\IP^1$ as well as properties of the limit
discussed in \cite{KKV}.
The derivation is similar as for the constant map contribution
$\int_{\overline{{\cal M}_g}} c^3_{g-1}=
\frac{|B_{2g} B_{2g-2}|}{2 g ( 2 g-2) (2g-2)!}$ in \cite{KKRS}.
The asymptotic in the magnetic expansion comes from the
occurrence of the $c=1$ string at the conifold \cite{Ghoshal:1995wm}.

We come now to the explicite iterative solutions in the genus. 
For example the recursive definition of $F^{(2)}$ is
\begin{equation}
{F}^{(2)}(\tau, \bar \tau)= {1\over 2} S {F}^{(1)}_{,2}+ {1\over
2} S ({F}^{(1)}_{,1})^2 + {5\over 24} S^3 ({F}^{(0)}_{,3})^2-
{1\over 8}S^2 {F}^{(0)}_{,4}- {1\over 2}S^2 {F}^{(1)}_{,1}
{F}^{(0)}_{,3}+ X c^{(2)}_0(u)\ , \label{F2}
\end{equation}
were $c^{(2)}_0(u)$ is the holomorphic ambiguity at genus two. This 
ambiguity must be invariant under $\Gamma(2)$, which implies that it
can be written in terms of $u$. Moreover regularity of $F^{(g)}$ at $u\rightarrow \infty $
and the leading pole behaviour at $u\rightarrow 1$ implies that it is of the form
\begin{equation}
X^{g-1} c^{(g)}_0(u)=u^{3-g}\sum_{i=1}^{2g-2} {A^{(g)}_i\over \Delta^i(u)}\ .
\label{swambiguity}
\end{equation}
Note that $A^{(g)}_i$ are undetermined constants and the right hand is a rational 
function of the $\Gamma(2)$ invariant function $u$.
Using (\ref{Caaa},\ref{anholF1}) and the standard formulas for 
the derivatives of $\theta_i,E_2$ 
we may write (\ref{F2}) with $h=(b+ 2 d)$ as
\begin{equation}
{F}^{(2)}(\tau,\bar \tau)= {X \over 3} \left(5 \hat E_2^3- 9 \hat E_2^2 h+
6 \hat E_2 (b^2+c d )- {h (16 b^2 +19 c d )\over 10}\right)\ ,
\label{swF2}
\end{equation}
an almost holomorphic modular function of $\Gamma(2)$. Here we determined the
ambiguity as follows. Using (\ref{limits},\ref{a-period}) we can expand (\ref{swF2})
in the electric and magnetic holomorphic limits. With the
leading behaviour (\ref{asym}) we found $A^{(2)}_1=
-{19\over 12960}$ and $A^{(2)}_2=-{2\over 405}$.
This yields
\begin{equation}
{\cal F}^{(2)}=-\frac{1}{240 \cdot 2\, a^2} - \frac{11}{2^{18}\,a^{10}} -
  \frac{117}{2^{22}\,a^{14}} -
  \frac{171201}{2^{34}\,a^{18}}+O({1\over a^{22}})
\end{equation} 
This series predicts all genus 2 instantons and checks with the
coefficients that appear in the literature \cite{Nekrasov:2002qd}\cite{KMT}. The
expansion in $a_D$ is obtained from (\ref{swF2}) using (\ref{limits},\ref{modtrans},\ref{ad-period})
\begin{equation}
{\cal F}^{D (2)}=-\frac{1}{240\, {\tilde a_D}^2} - 
\frac{\tilde a_D}{2^{13}} - \frac{13\,{\tilde a_D}^2}{2^{16}} - 
\frac{129\, {\tilde a_D}^3}{2^{17}5} + \
{O({\tilde a_D}^4)}
\label{fadII}
\end{equation}

Solving the recursion for genus 3 yields \cite{Klemm:1999gm}
\begin{equation}
\begin{array}{rl}
{F}^{(3)}=&
S{F}^{(2)}_{, 1}{F}^{(1)}_{, 1}
-{1\over 2}S^2{F}^{(2)}_{, 1}{F}^{(0)}_{, 3}
+ {1\over 2} S{F}^{(2)}_{, 2}
+ {1\over 6}S^3({F}^{(1)}_{, 1})^3 {F}^{(0)}_{, 3}
- {1\over 2}S^2{F}^{(1)}_{, 2}({F}^{(1)}_{, 1})^2 \cr &
- {1\over 2}S^4({F}^{(1)}_{, 1})^2({F}^{(0)}_{,3})^2
+ {1\over 4}S^3({F}^{(1)}_{, 1})^2{F}^{(0)}_{, 4}
+ S_2^3{F}^{(1)}_{, 2}{F}^{(1)}_{, 1}{F}^{(0)}_{, 3}
- {1\over 2}S^2{F}^{(1)}_{, 3}{F}^{(1)}_{, 1} \cr &
- {1\over 4}S^2({F}^{(1)}_{, 2})^2
+ {5\over 8}S^5{F}^{(1)}_{, 1}({F}^{(0)}_{, 3})^3
- {2\over 3}S^4{F}^{(1)}_{, 1}{F}^{(0)}_{, 4}{F}^{(0)}_{, 3}
- {5\over 8}S^4{F}^{(1)}_{, 2}({F}^{(0)}_{, 3})^2 \cr &
+{1\over 4}S^3{F}^{(1)}_{, 2}{F}^{(0)}_{, 4}
+{5\over 12} S^3{F}^{(1)}_{, 3}{F}^{(0)}_{, 3}
+ {1\over 8}S^3{F}^{(0)}_{, 5}{F}^{(1)}_{, 1}
- {1\over 8} S^2{F}^{(1)}_{, 4}
- {7\over 48} S^4{F}^{(0)}_{, 5}{F}^{(0)}_{, 3} \cr &
+ {25 \over 48}S^5{F}^{(0)}_{, 4}({F}^{(0)}_{, 3})^2
-{5\over 16}S^6 ({F}^{(0)}_{, 3})^4
- {1\over 12} S^4({F}^{(0)}_{, 4})^2
+{1\over 48} S^3{F}^{(0)}_{, 6}
+X^2 c^{(3)}_0(u)\ ,
\end{array}
\label{F3}
\end{equation}
and we determined the coefficients
$A^{(3)}_1=\frac{59}{2449440}$,
$A^{(3)}_2=\frac{14669}{3265920}$,
$A^{(3)}_3=\frac{4133}{204120}$ and
$A^{(3)}_4=\frac{5359}{306180}$. This yields the following almost complex
modular expression for ${F}^{(3)}$
\begin{equation}
\begin{array}{rl}
{ F}^{(3)}=& {4 X^2} \biggl( 5 \hat E_2^6- 25 \hat E_2^5 h + 40
\hat E_2^4 (2 b^2 + 5 cd)-{1\over 3} \hat E_2^3 h (529 b^2 + 559 c
d)+\cr & {\hat E_2^2\over 5} (1172 b^4 + 4060 b^2 c d + 1223
(cd)^2 )- \hat {E_2 \over 5} h (844 b^4 + 1685 b^2 c d + 310
(cd)^2)+\cr & {1\over 210} (10718 b^6 + 49596 b^4 c d + 44007 b^2
(cd)^2 + 944 (cd)^3) \Biggr)\ ,
\end{array}
\end{equation}
from which the electric
\begin{equation}
{\cal F}^{(3)}=\frac{1}{1008\cdot 2^3}\left(-\frac{1}{a^4} +
  \frac{441}{2^{12} a^{12}} +
  \frac{18459}{2^{16} a^{16}} +
  \frac{ 62106849 }{2^{28} a^{20}} +
  \frac{256368735}{2^{31} a^{24}}
 + O(\frac{1}{a^{28}})\right)
\end{equation}
and magnetic expansions
\begin{equation}
{\cal F}^{D(3)}=\frac{1}{1008}\frac{1}{{\tilde a_D}^4} - \frac{9\, \tilde a_D}{2^{20}} - 
\frac{143\, {\tilde a_D}^2} {2^{22}} - \frac{63827\, {\tilde a_D}^3}{2^{27} 7} + 
{O({\tilde a_D}^4)}
\label{fadIII}
\end{equation}
follow from (\ref{limits}). The number of terms in the modular
expressions of $F^{(g)}$ grow much slower then the number of graphs in the
holomorphic anomaly equation, because many graph contributions are
proportional to the same quasimodular form. For genus four, where the
holomorphic anomaly equation has 83 graphs we find
\begin{eqnarray}
{F}^{(4)} &=& 4X^3 \big\{
\frac{1150}{9}\hat{E}_2^9-985\hat{E}_2^8h+2\hat{E}_2^7(2399b^2+7001cd)
\nonumber \\&&
-\frac{14}{3}\hat{E}_2^6h(3761b^2+6125cd)+
\frac{3}{5}\hat{E}_2^5(85863b^4+363344b^2cd+240083(cd)^2)
\nonumber \\ &&
-\frac{1}{5}\hat{E}_2^4h(604469b^4+1677340b^2cd+547811(cd)^2)
\nonumber \\ &&
+\frac{2}{5}\hat{E}_2^3(531266b^6+2793615b^4cd+3285123b^2c^2d^2+447656(cd)^3))
\nonumber \\ &&
-\frac{2}{35}\hat{E}_2^2h(4430756b^6+16550337b^4cd+11925927b^2(cd)^2+889964(cd)^3)
\nonumber \\ &&
+\frac{1}{175}\hat{E}_2(31232428b^8+195274840b^6cd+329613819b^4(cd)^2+130729960b^2(cd)^3
\nonumber
\\ &&  +3566728(cd)^4)
-\frac{1}{1575}h(87826748b^8+423770948b^6cd \nonumber \\ &&
+511313601b^4(cd)^2+128098172b^2(cd)^3+4442006(cd)^4)
 \big\} \label{swF4}
\end{eqnarray}
with electric
\begin{equation}
{\cal F}^{(4)}=
 \frac -{1}{1440\cdot 2^5}\left(
  \frac{1}{a^6}+
  \frac{765}{2^{12} a^{14}}+
  \frac{126195}{2^{16} a^{18}} +
  \frac{1925006715}{2^{29} a^{22}} +
  \frac{14420664765}{2^{32} a^{26}} +
  {\cal O}(\frac{1}{a^{30}})\right)
\end{equation}
and magnetic expansion
\begin{equation}
{\cal F}^{D(4)}=-\frac{1}{1440{\tilde a_D}^6} + 
\frac{1125\ {\tilde a_D}}{2^{29}} - 
\frac{3915\, {\tilde a_D}^2}{2^{28}} + 
\frac{4786021\, {\tilde a_D}^3}{2^{35}} + {O({\tilde a_D}^4)} 
\label{fadIV}
\end{equation}

We derived expressions for $F^{(g)}$ in terms of modular forms up
to genus six and checked the large $a$ expansions against results
made available to us by Nakajima\footnote{These somewhat lengthy
expressions are available on request.}. Let us report here the
dual expansions, which are interesting as they correspond to
perturbations of the $c=1$ string at the selfdual radius by momentum operators 

\begin{equation}
{\cal F}^{ D(5)}=\frac{1}{1056\, {\tilde a_D}^8} - \frac{77175\, 
{\tilde a_D}}{2^{36}} - \frac{100971\, {\tilde a_D}^2}{2^{32}} 
- \frac{5142558213\, {\tilde a_D}^3}{2^{43}11} + 
{O({\tilde a_D^4})}\ , 
\label{fadV}
\end{equation}

\begin{equation}
{\cal F}^{D(6)}=-\frac{691}{327600\, 
{\tilde a_D}^{10}} - \frac{18753525 {\tilde a_D}}{2^{44}} - 
\frac{16908525{\tilde a_D}^2}{2^{40}} - \frac{672990085791 
{\tilde a_D}^3}{2^{49} 13} + {O({\tilde a_D^4})}\ .
\label{fadVI}
\end{equation}

\subsection{Fixing the ambiguity}
In (\ref{swsu2ambiguity}) all $c^{(g)}_k$  except for $c^{(g)}_0$, 
the holomorphic ambiguity, are determined by the recursion relations 
(\ref{F2},\ref{F3}) that follow from the holomorphic anomaly equation 
in terms of lower genus $F^{(g)}$. Genrally fixing the holomorphic ambiguity 
is a major problem in the B-model, which however the case at hand is 
completly solvable. The discriminant locus of the curve (\ref{curve1}) is at
$u=\infty$ and at $u=\pm 1$ where (\ref{curve1}) develops a node. 
At all other points in the moduli space of the
Seiberg-Witten geometry $F^{(g)}$ must be regular. As follows
from the global properties of the $\theta$ functions and $E_2$, 
regularity at $u=0$ restricts the form of  $c^{(g)}_k(\tau)$ to
\begin{equation}
c^{(g)}_k= h^{(1+(-1)^{g+k})\over 2} P_{d(g,k)}(b^2, c d),
\label{cgk}
\end{equation}
where $P_{d(g,k)}(b^2,c d)$ is an homogeneous polynomial in $b,c,d$ of
degree
\begin{equation}
d(g,k)=3(g-1)-k-{(1+(-1)^{g+k})\over 2}\ ,
\end{equation}
with ${d(g,k)\over 2}+1$ coefficients. In particular in the
ambiguity $c^{(g)}_0$ the number of unknown coefficients of this
polynomial grows with $\sim {3\over 2}g$ slower then the number
of the $A_i^{(g)}$ in (\ref{swambiguity}), which grows with $\sim 2g$.

Moreover the leading terms in (\ref{fadII},\ref{fadIII},\ref{fadIV},\ref{fadV},\ref{fadVI}) 
correspond to correlators of the cosmological
constant  operator of the $c=1$ matrix model in the genus g vacuum sector \cite{Ghoshal:1995wm}. 
Very important is the occurence of the {\sl  gap} in ${\cal F}^{(g) D}$, 
i.e. the absence  of  terms  $\frac{1}{{\tilde a_D}^k}$, $k=0,\ldots,2g-3$. 
The $2g-2$ gap conditions fix the $A^{(g)}_i$, $i=1,\ldots,2g-2$ constants and hence the ambiguity 
(\ref{swambiguity}). Together with the anomaly equation this 
provides a very efficient way to solve the theory completly.  
The asymptotic (\ref{asym}) and the particular form of (\ref{cgk}) 
are further consistency constraints confirming the gap property. 
A gap follows if there is a matrix model ${1\over N}$ expansion for the holomorphic 
topological string at a critical point, as e.g. at the orbifold point 
in the local ${\cal O}(-2,-2)\rightarrow \IP^1\times \IP^1$ model discussed \cite{AKMV}. 
In this case the measure integration  yields at each genus the negative power 
term in expansion parameter and the perturbative terms start with positive powers.
This particular behaviour of the magnetic expansion of the N=2 pure SU(2) model 
at the conifold is hence explainable by the proposed unitary matrix model \cite{DV2}.
As we saw above the model is solved by the gap property and the 
holomorphic anomaly. If one would like to employ ${1\over N}$ 
techniques in order to determine the weak coupling instanton expansion 
one would have to do exactly what we  have done in (\ref{swsu2ambiguity})  
namely to write the result globally. More generally the gap can be understood from the 
absence of correlators of the ground ring operators \cite{Witten:1991zd} in the $c=1$ string 
describing the limit of the toplogical string near the nodal
singularity (conifolds) of local models. Indeed we have checked that the gap 
occurs also at the conifold in the local ${\cal O}(-3)\rightarrow \IP^2$ geometry 
and fixes the ambiguity of this model. How this extends generally to 
singularities of local models and the modifications for singularities 
of global models will be discussed in \cite{workinprogress}.

If we absorb $X$ into genus expansion parameter $\lambda$  in
(\ref{swsu2ambiguity}), it becomes a sum of a quasi-modular form.
The simplest example of such an expansion, where the coefficients
are however modular forms of the full modular group $\Gamma_0$
appears in Hurwitz theory on $T^2$ \cite{Dijkgraaf}. As reviewed
there this leads directly to combinatorial problem that is solved
by free fermions and $Z=e^F$ can be written in a product form that
has been recognized as a generalized $\theta$-function product
form in \cite{Zagier}. More examples of such product forms are
provided by vertex algebras \cite{Borcherds,kontsevich1} and arise
in heterotic type II duality
\cite{Kawai,hosono,KKRS,Klemm:2005pd}. It would be interesting to
see whether the SW partition function is related to a generalized
$\Gamma(2)$ theta function.

\section{Dijkgraaf-Vafa conjecture} \label{DVsection1}
In \cite{DV1}, Dijkgraaf and Vafa proposed a remarkable relation
between B-model topological string on a non-compact Calabi-Yau geometry and
a matrix model.
\subsection{Dijkgraaf-Vafa transition and geometric engineering}
The $n$ cut matrix model is obtained by reducing  holomorphic
Chern-Simons theory on D5-branes wrapping $n$ $\IP^1$'s
in a modification of the geometry ${\cal O}(-2)\oplus
{\cal O}(0)\rightarrow \IP^1$.  The k-th $\IP^1$ is wrapped
by $N_k$ branes, $k=1,\ldots,n$. In the modified geometry
the location of the $\IP^1$'s in the originally flat ${\cal O}(0)$
$x$-direction is now fixed at the minima of a potential $W(x)$ of
degree $n+1$. E.g. the $n=1$ geometry  is the blown up conifold
${\cal O}(-1)\oplus {\cal O}(-1)\rightarrow \IP^1$.
The reduction yields a complex bosonic matrix model with
the matrix potential\footnote{Further generalization of
this conjecture to the case of Calabi-Yau geometry with ADE type
singularities can also be made~\cite{DV2}.} $W^{\prime}(x)$,
that needs as additional data the choice of contour for the
eigenvalue integration~\cite{DV1}.

The B-model geometry emerges after a transitions in which  the
$n$ $\IP^1$'s are shrunken and deformed to $S^3$'s.
It has a local description as a hypersurface in $\mathbb{C}^4$
\begin{equation}
vw=W^{\prime}(x)^2+f(x)+y^2
\label{eq:geometry}
\end{equation}
where $x,y,v,w$ are coordinates of $\mathbb{C}^4$, $f(x)$ is
polynomial of degree $n-1$ that splits the $n$ double zeros
of $W^{\prime}(x)^2$.

The latter geometry has been considered in \cite{CIV} to
geometrically engineer ${\cal N}=1$ four-dimensional
supersymmetric gauge theory. After the transition
the breaking to ${\cal N}=1$ is achieved by putting $N_k$ units of
Ramond flux  on the $S^3$'s and the topological string or
the matrix model calculates terms in the ${\cal N}=1$
effective gauge theory. In \cite{DV1} it is
already shown that the special geometry relation that
determine the tree level (genus zero) topological
string amplitude $F^{(0)}$ on the geometry (\ref{eq:geometry}) arise from the
planar diagrams of the corresponding matrix model. The planar loop
equation of the matrix model gives the spectral curve of the local
geometry and the effective superpotential in a large class of
$\mathcal{N}=1$ gauge theory can be computed exactly by the genus
zero amplitude of matrix model or topological string.
It is conjectured that higher genus topological B-model string
amplitudes should also be computed by higher genus diagrams in the
matrix model.

The meaning of the topological string amplitudes $F^{(g>0)}(S_i)$
in the effective theory is as follows. In $\mathcal{N}=2$ supergravity action
they determine the exact moduli dependence of the $F$-terms
\begin{equation} \label{F-term}
\int d^4x\int d^4\theta F^{(g)}(S_i)
(\mathcal{W}^{\alpha\beta}\mathcal{W}_{\alpha\beta})^g
\end{equation}
here $\mathcal{W}_{\alpha\beta}$ is the $\mathcal{N}=2$
graviphoton superfield and the $S_i$ are the $\mathcal{N}=2$
vector multiplets whose complex scalar field corresponds to the
moduli of Calabi-Yau manifold. After integrating over the
$\mathcal{N}=2$ superspace these terms become the coupling $R_+^2 F^{2g-2}_+$
of the self-dual part of the Ricci tensor $R_+$ to the selfdual
part of the graviphoton field strength $F_{+}$.

After breaking of $\mathcal{N}=2$ to $\mathcal{N}=1$ by fluxes the topological string
amplitudes $F^{(g>0)}(S_i)$ occur in the following two terms of the
$\mathcal{N}=1$ action~\cite{OV2, OV3, DGOVZ}
\begin{eqnarray} \label{N1}
&& \Gamma_1=g\int d^4x\int d^2\theta
\mathcal{W}_{\alpha\beta\gamma}\mathcal{W}^{\alpha\beta\gamma}(F_{\delta\xi}F^{\delta\xi})^{g-1}F^{(g)}(S_i),
\\ \label{N2} && \Gamma_2=\int d^4x\int d^2\theta
(F_{\alpha\beta}F^{\alpha\beta})^gN_i\frac{\partial
F^{(g)}(S_i)}{\partial S_i}\ .
\end{eqnarray}
Here $\mathcal{W}_{\alpha\beta\gamma}$ are now $\mathcal{N}=1$
gravitino multiplet, and $S_i$ are the $\mathcal{N}=1$ glueball
chiral superfields coming from the original $\mathcal{N}=2$ vector
multiplets \cite{OV2}. The graviphoton field $F_{\alpha\beta}$ can be treated
as background field in the $\mathcal{N}=1$ theory. In \cite{OV2} a
C-deformation is introduced to deform the anti-commutation
relation of gluino $\psi_{\alpha}$ to the followings
\begin{equation}
\{\psi_{\alpha},\psi_{\beta}\}=2F_{\alpha\beta}
\end{equation}
It is shown that the effect of turning on the graviphoton
background $F_{\alpha\beta}$ can be captured by this
C-deformation, and the $F^{(g)}$ in the second contribution
$\Gamma_2$ (\ref{N2}) are computed by matrix models at genus $g$.
We can also see that the first term $\Gamma_1$ in (\ref{N1})
contributes at genus $g=1$ even there is no graviphoton background
$F_{\alpha\beta}$. It is shown in \cite{OV3} that this genus one
contribution is also computed by matrix model genus one diagrams.
There are also some other types of gravitational corrections
besides (\ref{N1},\ref{N2}) of the form
$W_{\alpha\beta\gamma}W^{\alpha\beta\gamma}S^n$ from planar
diagrams, which become trivial after the extremization of the glue
ball superfield $S$ \cite{DGOVZ}. Our results confirm these very
interesting ideas in \cite{OV2,OV3,DGOVZ} by a first direct tests
of the connection between topological strings and matrix models at
higher genus without using the superspace techniques in the
effective gauge theory.

\subsection{The two cut geometry and the tree level and genus one amplitudes}

We consider now the case of a cubic potential $W(x)=\frac{m}{2}
x^2+\frac{g}{3}x^3$ in the Dijkgraaf-Vafa geometry
(\ref{eq:geometry}). The degree one polynomial
$f(x)=\mu_1 x+\mu_0$ splits the double zeros of
$W^{\prime}(x)^2$ at $x=a_1$ and $x=a_2$ to the four roots
$a_1^{\pm},a_2^{\pm}$ of the equation
\begin{equation}
W^{\prime}(x)^2+f(x)=0
\label{eq:lcy}
\end{equation}
We adopt the notation of \cite{KMT, CIV}, and change variable
$(a_1^{-},a_1^{+},a_2^{-},a_2^{+})\equiv (x_1,x_2, x_3,x_4)
\rightarrow (z_1,z_2, Q, I)$ where
\begin{eqnarray}
&&
z_1\equiv\frac{1}{4}(x_2-x_1)^2,~~~~z_2\equiv\frac{1}{4}(x_4-x_3)^2
\nonumber \\
&& Q\equiv \frac{1}{2}(x_1+x_2+x_3+x_4)=-\frac{m}{g} \nonumber \\
&&I\equiv\frac{1}{4}[(x_3+x_4)-(x_1+x_2)]^2=\left({m\over g}\right)^2-2(z_1-z_2)\ .
\end{eqnarray}
The local Calabi-Yau geometry (\ref{eq:lcy}) depends really only
two complex structure deformations. We will use $(z_1,z_2)$ or
below the A-periods $S_1$ and $S_2$ to parameterize them. The
dependence of the genus zero super- and the higher genus
potentials $F^{(g)}$ can be reconstructed from scaling laws and
dimensional considerations. In particular we frequently set
$g,m=1$.

The fundamental periods of the local geometry are
$$S_i=\frac{1}{2\pi i}\int_{a_i^{-}}^{a_i^{+}}\lambda, \qquad
\Pi_i=\frac{1}{2\pi i}\int_{a_i^{+}}^{\Lambda}\lambda\ ,$$
where $i=1,2$ and $\lambda=dx\sqrt{W^{\prime}(x)^2+f(x)}$ is a
meromorphic differential which emerges after integrating the
holomorphic Calabi-Yau (3,0)-form over an $S^2$ fibre direction
of the ${S^3}'$s in (\ref{eq:geometry}) \cite{CIV}.
In~\cite{CIV} the integrals where further calculated
perturbatively for small $z_i$. This limit corresponds to vanishing
${S^3}'s$ and is suitable for the perturbative matrix model expansion.
Solving the B-model and fixing its ambiguity requires a global understanding
of the complex moduli space in $(z_1,z_2)$. We therefore derive the
Picard-Fuchs equations and use them to explore the global properties
of the integrals $S_i$ and $\Pi_i$.

We find that derivatives of $\lambda$ w.r.t. $z_i$ up to second
order multiplied with suitable polynomials in $z_i$
combine to a total derivative, i.e. ${\cal L}_i\lambda=
\sum_{k+l\le 2} p^{(k,l)}({\underline z})
\partial^k_{z_1}\partial^l_{z_2}\lambda=\dd g_i(x,z_1,z_2)$. Naively
the differential ideal with this property is generated by three independent
differential operators ${\cal L}_i$.  However $\dd g_i(x,z_1,z_2)$ is a meromorphic
differential with non-vanishing residua, hence one cannot conclude from
the exactness that ${\cal L}_i\int_{\Gamma} \lambda=0$. For the
following two operators the residua vanish
\begin{equation}
\begin{array}{rl}
{\cal L}_1=&\ds{
2 z_1 (2 z_1 + 6 z_2 -1)  \partial_{z_1}^2 +
( 1 - 10 z_1 + 12 {z_1}^2 + 4 z_1 z_2 )\partial_{z_1}\partial_{z_2}
+ ( 3 - 2 z_1 - 6 z_2 ) \partial_{z_1}} \\ &+ (1 \leftrightarrow 2) \\[3 mm]
{\cal L}_2=&\ds{ (2 z_1 + 2 z_2 -1)[ -2 z_1  ( 1 - 4 (z_1+z_2) + 5
{z_1}^2 - 2 z_1 z_2 - 3 {z_2}^2 )  \partial_{z_1}^2  +( z_1 + z_2
)} \\ & \ds{( 1 - 8 z_1 + 6 {z_1}^2 - 6 z_1 z_2 )
\partial_{z_1}\partial_{z_2}]+ (z_1 ( 7 - 18 z_2 + 26 {z_2}^2 + 46
{z_1}^2 + {z_1}( 62 z_2 -39))} \\& \ds{ - 3 z_2 ( 1 - 3 z_2 + 2
z_2^2 ))) \partial_{z_1}- 3 ( 1 - 12 z_1 + 18 {z_1}^2 + 14 z_1 z_2
)+ (1 \leftrightarrow 2)\ .}
\end{array}
\label{eq:pf}
\end{equation}
These Picard-Fuchs operators annihilate the periods and fix their expansion up to linear combinations.
The discriminant of these differential operators has the following components
\begin{equation}
\begin{array}{rl}
C_1:& z_1=0, \quad C_2: \ z_2=0, \quad I: \ I=1-2 z_1 - 2 z_2=0,\\[2 mm]
J:&J=(x_1-x_2)(x_2-x_3)(x_2-x_4)(x_1-x_4) \\ [ 2mm]
&\phantom{J}= (1-3 z_1)^2-6 z_2+9z_2^2+14 z_1 z_2=0,
\label{eq:disc}
\end{array}
\end{equation}
whose schematic intersection after a suitable desingularisation of three order
tangencies is depicted in Fig. 1.
\begin{center}
\begin{figure}[htbp]
~~~~~~~~~~~~~~~~~~~~~~~~~\epsfig{file=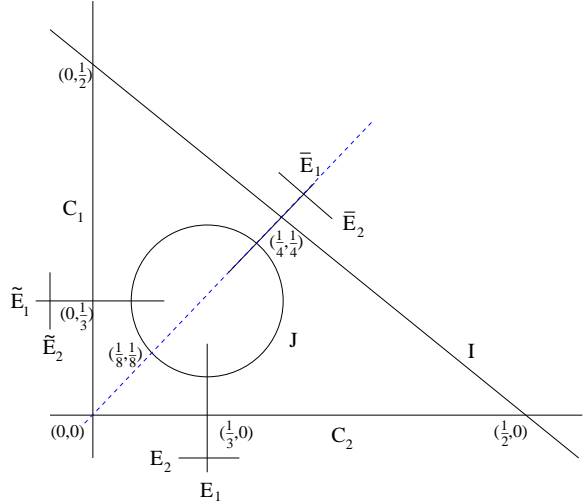,width=.5\textwidth}
\caption{\sl Divisors in the moduli space of B-model,
$(z_1,z_2)=(0,0)$ is the matrix model expansion point and along
the dash line one has enhanced $N=2$ SUSY.} \label{fig:2}
\end{figure}
\end{center}
\vskip -20pt
According to the Dijkgraaf-Vafa correspondence the periods $S_i$
are identified with the filling fractions $N_i$, with $\sum_{i=1}^n N_i=N$,
of eigenvalues in the large $N$ limit of the dual matrix model, and
it is shown that the special geometry relation and Picard-Fuchs equations are
reproduced in the planar limit of the matrix model~\cite{DV1}  and
the genus zero topological string amplitude $F^{(0)}$ follows from
integrating the special geometry relation ${\partial F^{(0)}\over \partial S_i}=\Pi_i$.
To analyze the exact effective superpotential $W_{eff}=2 \pi i
\sum_{i}(N_i \Pi_i+\alpha_i S_i)$, where $N_i$ and $\alpha_i$
are 3-form flux quanta through the $A_i$ and $B_i$ cycles respectively,
globally we need to the periods troughout the complex moduli space.
This is done in appendix A, where we also find that there is no point
where the periods degenerate quadratically in the logarithms.Which is
the signal of a large volume or maximal unipotent point in the
moduli space.

Also at one loop the conjecture holds\footnote{The one loop-test
in\cite{DST} and  the higher loops tests \cite{KMT} are made at
the $N=2$ point $S_1=-S_2$.}~\cite{KMT,DST}. The B-model expression for the
one loop free energy $F^{(1)}$ obtained by integrating (\ref{eq:anomalyrecursion}) using (\ref{special})
and fixing integration constants it turns out to be~\cite{KMT}
\begin{equation}
F^{(1)}={1\over 2} \log\left({\rm det}\left(\partial z_i\over \partial S_j\right)
(z_1 z_2)^{-{1\over 6}} I J^{2\over 3}\right) \ ,
\label{eq:genusone}
\end{equation}
where we obtain the periods $S_i$ in terms of complex
structure moduli $z_i$ and the corresponding inverse series
\begin{eqnarray}
&&
S_1(z_1,z_2,g)=\frac{1}{4}z_1-\frac{g^2}{8}z_1(2z_1+3z_2)-\frac{g^4}{32}z_1(4z_1^2+13z_1z_2+9z_2^2)+\mathcal{O}(g^6)
\nonumber \\
&& S_2(z_1,z_2,g)=-S_1(z_2,z_1,g) \nonumber \\
&&
z_1(S_1,S_2,g)=4S_1+8g^2S_1(2S_1-3S_2)+8g^4(20S_1^2-67S_1S_2+39S_2^2)+\mathcal{O}(g^6)
\nonumber \\
&& z_2(S_1,S_2,g)=z_1(-S_2,-S_1,g)
\end{eqnarray}
from the two power series solutions to (\ref{eq:pf}) at
$(z_1,z_2)=(0,0)$. Identifying $S_i$ with the filling fractions
$N_i$ we get the genus one contributions to (\ref{eq:fmatrix}).
Note that the integration constants $c_i$ in $(z_1 z_2)^{c_1}
I^{c_2} J^{c_3}$, which fix the behaviour of $F^{(1)}$ at the
discriminant components are global data do not depend on the base
point $(z_1,z_2)=(0,0)$ or the holomorphic limit $\bar S_{\bar
\imath}\rightarrow 0$ taken at this base point to obtain the
matrix model expansion. The coefficient $c_1=-{1\over 6}$ at the
conifold (shrinking $S^3$) is a universal property of the
topological B-model.

To solve the B-model recursion we need the genus zero three point functions,
which are rational functions in complex structure moduli.
\begin{equation}
C_{z_1z_1z_1}= \frac{1-(6z_1+5z_2)g^2+3(3z_1^2+3z_1z_2+2z_2^2)g^4}{16g^4z_1I^2},\quad
C_{z_1z_1z_2}=\frac{1-(3z_1+5z_2)g^2}{16g^2I^2}\ .
\end{equation}
The three point functions $C_{z_iz_jz_k}$ are symmetric in $ijk$
and from symmetry consideration follows $C_{z_2z_2z_2}=C_{z_1z_1z_1}(z_1\leftrightarrow z_2)$ as well as
$C_{z_2z_1z_1}=C_{z_1z_1z_1}(z_1\leftrightarrow z_2)$.

The corresponding matrix model is a Hermitian matrix model with
the cubic potential $W(\Phi)=\frac{1}{2}\Phi^2+\frac{g}{3}\Phi^3$
for a rank $N$ Hermitian matrix $\Phi$. The partition function and
free energy $F$ of the model are
\begin{equation}
Z=e^F=\frac{1}{Vol(U(N))}\int D\Phi~e^{-W(\Phi)}
\end{equation}

In the large $N$ limit the eigenvalues distribute around the two
critical points $0$ and $-\frac{1}{g}$ of the potential and form
two cuts. We consider the metastable vacuum where with $N_1$
eigenvalues at $0$ and $N_2$ eigenvalues at $-\frac{1}{g}$. This
is a two-cut solution of the matrix model with $N_1$ and $N_2$
fixed and subject to the condition $N_1+N_2=N$. In the large $N$
limit the free energy of the matrix model has genus expansion in
$1/N^2$, and at each genus there is a perturbative expansion by
the t'Hooft coupling constant $gN$. Dijkgraaf and Vafa conjecture
that the free energy of the matrix model at each genus is matched
to the topological string amplitudes on the local Calabi-Yau
geometry (\ref{eq:geometry}), by identifying of the periods $S_i$ in
the geometry with the eigenvalue filling fractions $N_i$. In
Appendix \ref{matrixmodel} we review more details of the matrix
model calculations of the free energy.

\section{The holomorphic anomaly equations} \label{DVsection2}
The key to the solution of the topological B-model are the
holomorphic anomaly equations. To solve them recursively one needs
in general to derive three types of propagators $S^{ij}$,
$S^{i\phi}$ and $S^{\phi\phi}$~\cite{BCOV}. In local geometries
$S^{i\phi}$ and $S^{\phi\phi}$ can be gauged to
zero~\cite{Klemm:1999gm} and the derivation of the propagators in
the multimoduli case is discussed in \cite{BCOV,KKRS,KM}.

\subsection{The holomorphic anomaly recursions}
The geometry on the complex structure moduli space $z_i$ of
Calabi-Yau is a special Kahler geometry. Its metric, connection
and curvature are determined by the Kahler potential $K$ by the
well-known formula,
$G_{i\bar{j}}=\partial_{i}\bar{\partial}_{\bar{j}}K$,
$\Gamma^{i}_{lm}=-G^{i\bar{k}}\partial_lG_{\bar{k}m}$ and
$R^k_{i\bar{j}l}=-\bar{\partial}_{\bar{j}}\Gamma^k_{il}$. They
have a well-known special geometry relation with the three point
Yukawa coupling $C_{ijk}=D_iD_jD_k F^{(0)}$, which comes from the
$tt^{*}$ equation \cite{CV} and can be thought of as the
holomorphic anomaly equation at genus zero
\begin{equation}
R^{k}_{i\bar{j}l}=G_{i\bar{j}}\delta^k_l+G_{k\bar{j}}\delta^k_i-C_{ilm}\bar{C}^{km}_{\bar{j}}
\label{special}
\end{equation}
At genus one and higher genus the topological string amplitudes
has a holomorphic anomaly, and the anti-holomorphic dependence of
the genus $g$ free energy is related to lower genus free energy by
the holomorphic anomaly equation \cite{BCOV}
\begin{equation}
\begin{array}{rl}
\partial_i\bar{\partial}_{\bar{j}}F^{(1)}&=\ \ds{
\frac{1}{2}C_{ikl} {\bar C}_{\bar \jmath}^{kl} -(\frac{\chi}{24}-1)G_{i\bar{j}}}\\ [2 mm]
\bar{\partial}_{\bar{i}}F^{(g)}&=\ \ds{\frac{1}{2}  {\bar C}_{\bar \imath}^{jk}
(D_jD_kF^{(g-1)}+\sum_{r=1}^{g-1}D_jF^{(r)}D_kF^{(g-r)})
\qquad g\geq 2}\ .
\end{array}
\label{eq:anomalyrecursion}
\end{equation}
Here the ${\bar C}_{\bar \imath}^{jk}=e^{2K} {\bar C}_{\bar \imath
\bar \jmath \bar k} G^{\bar \jmath j}G^{\bar k k}$ define the
propagators  $S^{ij}$ as
$\bar{\partial}_{\bar{i}}S^{jk}=\bar{C}_{\bar{i}}^{jk}$. The
holomorphic equation can be integrated and represented as graphic
Feynman rules to give the higher genus free energy in terms of
lower genus up to a holomorphic ambiguity. The propagators can be
solved by integrating its defining relation and use the special
geometry relation (\ref{special}). One finds
\begin{equation} \label{propagator}
S^{ij}C_{jkl}=\delta^i_l\partial_kK+\delta^i_k\partial_lK+\Gamma^i_{kl}+f^i_{kl}
\end{equation}
here $f^i_{kl}$ are ambiguous integration constants, and they are
meromorphic rational functions of the complex structure moduli
$z_i$ with poles at discriminant points of the moduli space.
Suppose there are $n$ complex structure moduli, then there are
$\frac{1}{2}n^2(n+1)$ equations for $\frac{1}{2}n(n+1)$
propagators $S^{ij}$. In the case of one modulus, the number of
equations and propagators are the same, so the meromorphic
functions $f^i_{jk}$ can be just set to zero. However, in the
multi-moduli case we consider, the equations over determine the
propagators, so we have to choose the ambiguity $f^i_{jk}$
properly to satisfy some constrains and ensure we can solve for
the propagators.

There are certain simplification in B-model calculations for the
case of non-compact local Calabi-Yau manifold. In this case there
is a choice of gauge such that the Kahler potential $K$ and metric
over the moduli space $G_{S_k\bar{S}_{\bar{j}}}$ in the $S_i$
coordinates is a constant in holomorphic limit, and the dilaton
component in the propagators vanish.  So in the holomorphic limit
$\bar{S}_{\bar{i}}\rightarrow 0$ the connection vanishes and
covariant derivative in the $S_i$ coordinates is just ordinary
derivative. This also makes the topological string amplitudes
entirely independent of quantities such as Euler number, Chern
classes of the Calabi-Yau, which will need to be regularized in
the non-compact case. In this case, the metric and the connection
in the $z_i$ coordinates are
\begin{eqnarray}
&&
G_{z_i\bar{z}_{\bar{j}}}=\frac{\partial{S_k}}{\partial{z_i}}C_{k{\bar{j}}}
\nonumber \\ &&
\Gamma^{z_i}_{z_jz_k}=-\frac{\partial{z_i}}{\partial{S_l}}\frac{\partial^2
S_l}{\partial z_j\partial z_k}
\end{eqnarray}
where $C_{k{\bar{j}}}$ are constant in the holomorphic limit.

\subsection{Propagators and Dijkgraaf-Vafa conjecture at higher genus}
The geometry we consider has two complex structure moduli. In
order to solve the propagators, the holomorphic ambiguity
$f^i_{jk}$ have to satisfy 3 constrain equations by eliminating
propagators $S^{ij}$ in (\ref{propagator}). These constrains
equations are rational functions of the complex structure moduli
$z_i$, \footnote{We note that while genus zero three point
functions are rational functions of the complex structure moduli
$z_i$, the connections $\Gamma^{i}_{jk}$ are generally not
rational functions. They combine to give rise to rational
equations for holomorphic ambiguity $f^i_{jk}$.} and we are able
to find a rational solution for the $f^{i}_{jk}$
\begin{equation}
\begin{array}{rl}
\label{ambiguity1}
f^1_{11}=&-\Big[6-(49z_1+48z_2)g^2+(163z_1^2+219z_1z_2+126z_2^2)g^4
\\  & +(210z_1^3+304z_1^2z_2+242z_1z_2^2+108z_2^3)g^6)\Big]
/(20g^6z_1  I^2 J) \\
f^1_{12}=&-\Big[29-(79z_1+157z_2)g^2+(10z_1^2+260z_1z_2+210z_2^2)g^4]/(20g^4 I^2 J)
\\
f^1_{22}=&\Big[7-(55z_1+68z_2)g^2+(142z_1^2+315z_1z_2+219z_2^2)g^4 \\
&-(120z_1^3+338z_1^2z_2+492z_1z_2^2+234z_2^3)g^6\Big]/(20g^6z_2 I^2 J). \\
\end{array}
\end{equation}
By definition $f^i_{jk}=f_{kj}$ and  the $(z_1\leftrightarrow
z_2)$ symmetry determines the other $f^i_{jk}$, e.g.
$f^2_{22}=f^1_{11}(z_1\leftrightarrow z_2)$ e.t.c. Note that the
discriminant factors in (\ref{eq:disc}) should be the only
singularities appearing in the denominator for the ansatz of
holomorphic ambiguities and in fact all appear.

The holomorphic anomaly equation at genus one can be integrated to
give the Ray-Singer torsion of the target manifold. The genus one
free energy can also be expressed in terms of genus zero three
point functions and the propopogators as follows
\begin{equation} \label{consistency}
\partial_i F^{(1)}=\frac{1}{2}S^{jk}C_{ijk}+\partial_i\sum_{r}a_r\log
(\Delta_r)
\end{equation}
here $a_r$ are constants and $\Delta_r$ are various discriminants
of the local geometry. The holomorphic ambiguity $f^i_{jk}$ should
give a solution of the propagators $S^{ij}$ that satisfies the
above consistency check (\ref{consistency}). We have chosen our
ansatz of the holomorphic ambiguity (\ref{ambiguity1}) that
satisfies the (\ref{consistency}) with the constants $a_1=1/15$,
$a_2=2/15$
\begin{equation}
\partial_i
F^{(1)}=\frac{1}{2}S^{jk}C_{ijk}+\partial_i(\frac{1}{15}\log(z_1z_2)+\frac{2}{15}\log
(\Delta_2))
\end{equation}
This choice of ansatz is convenient in the sense that it leads to
the correct leading behavior in genus two, so it is easier for us
to fix the genus two holomorphic ambiguity there.

In local geometry the genus two topological free energy can be
integrated from the holomorphic anomaly equation. It is
\begin{eqnarray}
F^{(2)} &=& \label{F2a}
-\frac{1}{8}S^{ij}S^{kl}F^{(0)}_{ijkl}+\frac{1}{2}S^{ij}F^{(1)}_{ij}
-\frac{1}{2}S^{ij}S^{kl}F^{(1)}_iF^{(0)}_{jkl}+\frac{1}{2}S^{ij}F^{(1)}_iF^{(1)}_j
\\ \nonumber &&
+\frac{1}{12}S^{ij}S^{kl}S^{mn}F^{(0)}_{ikm}F^{(0)}_{jln}+\frac{1}{8}S^{ij}S^{kl}S^{mn}F^{(0)}_{ijk}F^{(0)}_{lmn}+f^{(2)}
\end{eqnarray}
We fix the  genus two holomorphic ambiguity $f^{(2)}$ with some
initial data from matrix model calculations

\begin{equation}
\begin{array}{rl}
f^{(2)}=&
-\Big[1253-10503(z_1+z_2)g^2+27(1081z_1^2+950z_1 z_2+1081z_2^2)g^4 \\ [2 mm]
        &-26865(z_1+z_2)(z_1-z_2)^2g^6]/(9000g^4z_1 z_2 J^2),
\label{ambiguity2}
\end{array}
\end{equation}

Once the holomorphic ambiguity is fixed, we can compute the genus
two free energy to very high order using (\ref{F2a}). The
topological string approach is much more advantageous than direct
matrix model calculations where it is hard to compute to free
energy at higher orders (see Appendix \ref{matrixmodel} for more
details). After some extensive computer running time, we are able
to make many checks of the topological string predictions
(\ref{F2results}) below for the genus two free energy.
\begin{eqnarray} \label{F2results}
 F^{(2)} &=& -\frac{1}{240}(\frac{1}{{{N_1}}^2}+\frac{1}{{{N_2}}^2})+( \frac{35}{6}{N_1}-\frac{35}{6}{N_2} )g^6
 \nonumber \\ &+&
  ( 338{{N_1}}^2 - 1632{N_1}{N_2} + 338{{N_2}}^2 ) g^8
 + ( \frac{66132}{5}{{N_1}}^3 -
120880{{N_1}}^2{N_2}+\cdots )g^{10}  \nonumber \\ &+&
(\frac{1305280}{3}{{N_1}}^4 - \frac{18059582}{3}{{N_1}}^3{N_2}
+ 11824166{{N_1}}^2{{N_2}}^2 - \cdots) g^{12} \nonumber \\
&+&
    ( 12963696{{N_1}}^5 - 244438427{{N_1}}^4{N_2}  + 745362156{{N_1}}^3{{N_2}}^2 -
     \cdots ) g^{14} \nonumber \\ &+&
  ( 362264064{{N_1}}^6 - \frac{35002369227}{4}{{N_1}}^5{N_2} + \frac{148144671957}{4}{{N_1}}^4{{N_2}}^2
  \nonumber \\ && -
     57597548553{{N_1}}^3{{N_2}}^3 + \cdots ) g^{16} \nonumber \\
     &+&
  ( \frac{29035470208}{3}{{N_1}}^7 - \frac{862430780350}{3}{{N_1}}^6{N_2}  +
     1580964252892{{N_1}}^5{{N_2}}^2 \nonumber \\ &&  - \frac{10256675032550}{3}{{N_1}}^4{{N_2}}^3 + \cdots ) g^{18}
     \nonumber \\ &+&
  ( \frac{1250634104832}{5}{{N_1}}^8 - \frac{44363662176978}{5}{{N_1}}^7{N_2} +
     \frac{303466060570354}{5}{{N_1}}^6{{N_2}}^2 \nonumber \\ &&
     - \frac{854152004682126}{5}{{N_1}}^5{{N_2}}^3 +
     237637137780236{{N_1}}^4{{N_2}}^4 -\cdots ) g^{20}
\end{eqnarray}

The main difficulty in the B-model calculations is to fix the
holomorphic ambiguities at each genus. Also the Feynman rules that
solve the holomorphic equations quickly become very complicated.
Here we push the calculations only to genus two in our
calculations, since there are less new conceptual issues beyond
that.

\section{Conclusion}

The B-model iteration in the genus bears some resemblance to the
procedure compute higher genus free energy and resolvent of the
matrix models for one-cut solution in \cite{ACKM} and generalized
to multi-cut solution in \cite{Akemann,Chekhov:2005rr,Chekhov:2006rq},
where the iteration  equation is obtained by doing $1/N$ expansion in the loop
equations, and looks similar to the holomorphic anomaly equation
in topological strings.

From the hermitian matrix model point of view the
anti-holomorphicity is very unnatural. The holomorphic anomaly
equations were re-interpreted in \cite{wqbi} as infinitesimal
manifestation of the fact that the topological string partition
function transforms as a wave function under change of
polarisation in the middle cohomology of the target space. Using
this picture the failure of  holomorphicity can be traded against
a failure of modularity with a similar
iteration~\cite{ABK}, which makes the connection more
naturally.

It would be very interesting to compare this latter iteration
with the iterations \cite{Chekhov:2005rr,Chekhov:2006rq} in detail,
since this can in principle fix the holomorphic ambiguity in
B-model calculations. Fixing the holomorphic ambiguity systematically
is one of the main difficulties for topological string calculations in many other models.
We hope further studies will clarify these issues
and provide valuable lessons in fixing holomorphic ambiguity in
more general models.

\vspace{0.2in} {\leftline {\bf Acknowledgments:}}

A.K. likes to thank M.~Aganagic, V.~Bouchard, M.~Mari\~no, S.~Theisen and D.~Zagier 
for discussions, Nakajima for providing expressions for the higher instanton 
numbers and M.~Mantone for a correspondence. M.H. would like to thank Simons 
Workshop on Mathematics and Physics 2005, Hangzhou 2005 Winter Workshop
on String Theory, MSRI at Berkeley for hospitality during
parts of the work. Our work is
supported by DOE-FG02-95ER40896.

\appendix

\section{Moduli space and monodromy of the two cut matrix model}
\label{monodromy}

\subsection{Compactification of the moduli space and local expansions}
The aim of this section is to obtain the periods everywhere
in the moduli space and to determine the monodromies. For the
compactification of the moduli space we use the projective space
$\mathbb{C}\mathbb{P}^2$ with homogeneous coordinates
$(\tilde{z}_1,\tilde{z}_2,\tilde{z}_3)$ and identify the
$\tilde{z}_3\neq 0$ patch with
\begin{equation} \label{compact}
z_1=\frac{\tilde{z}_1}{\tilde{z}_3},~~z_2=\frac{\tilde{z}_2}{\tilde{z}_3}\ .
\end{equation}
In addition to the divisors listed in (\ref{eq:disc}) we get now a
$\mathbb{C}\mathbb{P}^1$ divisor at infinity, at which the periods
turn out to be non-singular. We calculated the local expansion
near all normal crossing divisors and determined the local
monodromy.  By analytic continuation we determined the global
mondromy. One remarkable aspect of the geometry is that there is
no point in the moduli space where at least one of the periods
degenerates with double logarithm, which would correspond to the
normal large complex structure point of a local geometry at which
the mirror expansion in the large K\"ahler coordinates leads to a
convergent instanton sum.

Suppose we expand the Picard-Fuchs equation around a common point
of two singular divisors $\Delta_1(z_1,z_2)=0$ and
$\Delta_2(z_1,z_2)=0$. In order to find complete solutions of the
Picard-Fuchs equation, one must choose a good local coordinate
around these singular points. The technique for choosing good
local coordinates is quite standard in algebraic geometry. For our
two parameter model, there are two possible cases:
\begin{itemize}
\item $\det(\frac{\partial\Delta_i}{\partial z_j})\neq 0$, then the
point $\Delta_1(z_1,z_2)=\Delta_2(z_1,z_2)=0$ is called the point
of normal intersection of divisor $\Delta_1=0$ and $\Delta_2=0$.
In this case a choice of good local coordinates is simply
$(\Delta_1,\Delta_2)$.

\item $\det(\frac{\partial\Delta_i}{\partial z_j})=0$, then this
is called a point of tangency of divisor $\Delta_1=0$ and
$\Delta_2=0$. In this case one will not be able to find all
solutions around the point of tangency with the choice of local
coordinates $(\Delta_1,\Delta_2)$. We will encounter a very common
situation in which the divisors have the following form
\begin{eqnarray}
\Delta_1&=&a^2+bc \nonumber \\
\Delta_2&=&b
\end{eqnarray}
here $a$, $b$ and $c$ are degree one polynomial of complex
structure moduli $z_1$ and $z_2$. The standard technique in
algebraic geometry is to introduce two exceptional divisors to
resolve the point of tangency. It turns out that a choice of good
local coordinate in this case is $(a, \frac{b}{a^2})$. In our
analysis we will follow the standard procedure and use this good
local coordinates.

\end{itemize}

We list the asymptotic solutions of Picard-Fuchs equations and
their monodromy at various singular points of the divisors
(\ref{eq:disc}) in the moduli space. Some of
the singular points can be obtained by exchanging $z_1$ and $z_2$,
and we only list once these symmetric singular points.
\begin{enumerate}
\item $C_1\cap C_2$. his is the matrix model point we have tested
Dijkgraaf-Vafa conjecture at higher genus. Th good choice of local
coordinates is simply $(z_1,z_2)$. For completeness we list the
periods
\begin{eqnarray}
f_1&=&\frac{z_1}{4}-\frac{1}{8}z_1(2z_1+3z_2)+\cdots \nonumber \\
f_2&=&-\frac{z_2}{4}+\frac{1}{8}z_2(3z_1+2z_2)+\cdots \nonumber \\
f_3&=&
f_1\log(z_1)+\frac{1}{12}+\frac{z_1}{8}-\frac{1}{16}(-4z_1^2+z_1z_2+5z_2^2)+\cdots
\nonumber \\
f_4&=&
f_2\log(z_2)-\frac{1}{12}-\frac{z_2}{8}-\frac{1}{16}(5z_1^2+z_1z_2-4z_2^2)+\cdots
\end{eqnarray}

\item $C_1\cap I$. The intersection point is at $(z_1,z_2)=(0,\frac{1}{2})$,
and the choice of good local coordinates is
$(x_1,x_2)=(z_1,1-2z_1-2z_2)$. The asymptotic solutions for
periods are
\begin{eqnarray}
f_1&=&\sqrt{x_2}(1-4x_1-x_2) \nonumber \\  f_2&=&
\sqrt{x_2}(x_1+\frac{x_2}{12}-(2x_1^2-\frac{5}{3}x_1x_2-\frac{x_2^2}{8})+\cdots)
\nonumber \\
f_3&=&x_1(1-\frac{5}{4}x_1-\frac{3}{2}x_2+\cdots) \nonumber \\
f_4&=&
f_3\log(x_1)+\frac{2}{3}+x_2+(\frac{7}{8}x_1^2-3x_1x_2-\frac{13}{12}x_2^2)+\cdots
\end{eqnarray}

\item $C_1\cap J$. This is a point of tangency of the two divisors at
$(z_1,z_2)=(0,\frac{1}{3})$. The singular factor $J$ can be
written as
\begin{equation}
J=1-6z_1-6z_2+9z_1^2+14z_1z_2+9z_2^2=9(z_2-\frac{1}{3})^2+z_1(-6+14z_2+9z_1)
\end{equation}
Following our discussion above we see a good choice coordinates is
$(x_1,x_2)=(\frac{z_1}{(z_2-\frac{1}{3})^2},\frac{1}{3}-z_2)$.
This is the local coordinates around the intersection of the blow
up divisor with the divisor $C_1$. The asymptotic expansion for
the periods are
\begin{eqnarray}
f_1 &=&
x_1x_2^{\frac{5}{2}}(1-\frac{1}{216}x_1-\frac{23}{72}x_1x_2-\frac{5}{46656}x_2^2+\cdots) \nonumber \\
f_2&=& x_2^2(1+\frac{4}{9}x_1+4x_2+\cdots) \nonumber \\
f_3 &=& 1+81x_2^3-108x_1x_2^3-\frac{729}{8}x_2^4+\cdots \nonumber
\\
f_4&=&
f_1\log(x_1)+x_2^{\frac{5}{2}}(\frac{36}{5}-\frac{108}{7}x_2+\cdots)
\end{eqnarray}
We can also solve the Picard-Fuchs equations with the local
coordinates around the intersection of the blow up divisor with
the singular divisor $J$. This will be useful later on when we try
to match the basis and derive the monodromy of the divisor $J$.
The good choice of local coordinates around this point is
\begin{eqnarray} x_1
=\frac{z_1(6-14z_2-9z_1)}{9(z_2-\frac{1}{3})^2}-1,~~~x_2=\frac{1}{3}-z_2
\end{eqnarray}
We find the asymptotic expansion for the periods with this
coordinates
\begin{eqnarray}
f_1 &=&
x_1^2x_2^{\frac{5}{2}}(1-\frac{5}{16}x_1+\frac{33}{8}x_2+\cdots)
\nonumber \\
f_2 &=& x_2^2(1+\frac{3}{4}x_1-\frac{43}{2}x_2+\cdots) \nonumber
\\
f_3 &=& 1-648x_2^3+\cdots \nonumber \\
f_4 &=&
f_1\log(x_1)+x_2^{\frac{5}{2}}(\frac{512}{15}+32x_1-\frac{3008}{7}x_2+\cdots)
\end{eqnarray}

\item $I\cap J$. This is a point of tangency between the
divisors at $(z_1,z_2)=(\frac{1}{4},\frac{1}{4})$. We write the
singular factor $J$ as
\begin{equation}
J=(z_1-z_2)^2+(1-2z_1-2z_2)(1-4z_1-4z_2)
\end{equation}
A good choice of local coordinates is
$(x_1,x_2)=((z_1-z_2)^2,\frac{1-2z_2-2z_2}{(z_1-z_2)^2})$. The
asymptotic solutions for the periods are
\begin{eqnarray}
f_1&=&
x_1(1+\frac{3}{8}x_1+x_2+\frac{35}{64}x_1^2+\frac{1}{4}x_1x_2+\cdots)
\nonumber \\
f_2 &=& x_1\sqrt{x_2} \nonumber \\
f_3 &=& f_1\log(x_1)
+\frac{4}{3}+\frac{23}{16}x_1^2+3x_1x_2+\cdots \nonumber \\
f_4 &=&
f_2\log(x_1)+x_1x_2^{\frac{3}{2}}(\frac{1}{3}+\frac{1}{30}x_2+\cdots)
\end{eqnarray}

\item $C_1\cap C_{\infty}$. In the homogeneous coordinate
$(\tilde{z}_1,\tilde{z}_2,\tilde{z}_3)$ the divisor $C_1$ is
$\tilde{z}_1=0$, so the good local coordinates is
$(x_1,x_2)=(\tilde{z}_1,\tilde{z}_3)$. Since at the intersection
$\tilde{z}_2\neq 0$, we can choose $\tilde{z}_2=1$ and use the
relation (\ref{compact}) to find
$(x_1,x_2)=(\frac{z_1}{z_2},\frac{1}{z_2})$. The asymptotic
solutions for the periods are
\begin{eqnarray}
f_1&=&
x_1x_2^{-\frac{3}{2}}(1-\frac{23}{72}x_1-\frac{1}{6}x_2+\cdots)
\nonumber \\
f_2&=&
x_2^{-\frac{3}{2}}(1-\frac{1}{4}x_2-\frac{67}{144}x_1^2+\frac{7}{24}x_1x_2
-\frac{1}{32}x_2^2+\cdots) \nonumber \\
f_3&=& f_1\log(x_1)+
x_2^{-\frac{3}{2}}(x_2-\frac{65}{144}x_1^2+\frac{1}{2}x_1x_2-\frac{1}{8}x_2^2+\cdots)
\nonumber \\
f_4 &=&(f_1-2f_2)\log(x_2)+x_2^{-\frac{3}{2}}
(\frac{2}{3}x_2-\frac{211}{108}x_1^2+\frac{25}{18}x_1x_2-\frac{1}{6}x_2^2+\cdots)
\nonumber
\end{eqnarray}

\item $I\cap C_{\infty}$. In the homogeneous coordinates
$\tilde{z}_1$, $\tilde{z}_2$ and $\tilde{z}_3$, the divisor $I$ is
$\tilde{z}_3-2\tilde{z}_1-2\tilde{z}_2=0$. At the intersection
with $I\cap C_{\infty}$ the coordinate $\tilde{z}_2\neq 0$, we can
choose $\tilde{z}_2=1$ and use the relation (\ref{compact}) to
find the good local coordinates
\begin{eqnarray}
(x_1,x_2)=(\tilde{z}_3-2\tilde{z}_1-2\tilde{z}_2,\tilde{z}_3)=(\frac{1-2z_1-2z_2}{z_2},\frac{1}{z_2})
\end{eqnarray}
The asymptotic solutions for the periods are
\begin{eqnarray}
f_1&=& \sqrt{x_1}x_2^{-\frac{3}{2}}(1+\frac{x_1-x_2}{4}) \nonumber
\\
f_2&=&
x_2^{-\frac{3}{2}}(1+\frac{3}{4}x_1-\frac{5}{64}x_1^2+\frac{3}{32}x_1x_2
-\frac{3}{64}x_2^2+\cdots) \nonumber \\
f_3&=& x_2^{-\frac{3}{2}}(x_1+x_2)(1+\frac{x_1-x_2}{8}+\cdots)
\nonumber \\
f_4&=&
f_1\log(x_2)+\sqrt{x_1}x_2^{-\frac{3}{2}}(-\frac{1}{4}x_1+\frac{1}{4}x_2
+\frac{13}{480}x_1^2+\frac{1}{48}x_1x_2-\frac{1}{32}x_2^2+\cdots)
\nonumber
\end{eqnarray}

\item $J \cap C_{\infty}$. In the homogeneous coordinates
$\tilde{z}_1$, $\tilde{z}_2$ and $\tilde{z}_3$, the divisor $J$ is
$\tilde{z}_3^2-6\tilde{z}_1\tilde{z}_3-6\tilde{z}_2\tilde{z}_3+9\tilde{z}_1^2+14\tilde{z}_1\tilde{z}_2+9\tilde{z}_2^2=0$.
At the intersection with $J \cap C_{\infty}$ the coordinate
$\tilde{z}_2\neq 0$, we can choose $\tilde{z}_2=1$ and use the
relation (\ref{compact}) to find the good local coordinates
\begin{eqnarray}
(x_1,x_2)&=&(\tilde{z}_3^2-6\tilde{z}_1\tilde{z}_3-6\tilde{z}_2\tilde{z}_3+9\tilde{z}_1^2+14\tilde{z}_1\tilde{z}_2+9\tilde{z}_2^2
,\tilde{z}_3) \nonumber \\
&=&(\frac{1-6z_1-6z_2+9z_1^2+14z_1z_2+9z_2^2}{z_2^2},\frac{1}{z_2})
\end{eqnarray}
The asymptotic solutions for the periods are
\begin{eqnarray}
f_1 &=&
x_1^2x_2^{-\frac{3}{2}}(1+\frac{103-\sqrt{2}i}{576}x_1+\frac{8+\sqrt{2}i}{16}x_2+\cdots)
\nonumber \\
 f_2 &=&
x_2^{-\frac{3}{2}}(1+\frac{-5-\sqrt{2}i}{64}x_1+\frac{-5-\sqrt{2}i}{16}x_2+\cdots)
\nonumber \\
f_3 &=& (f_1-\frac{8192(43+13\sqrt{2}i)}{2187}f_2)\log(x_2)
\nonumber \\ &+&
x_2^{-\frac{3}{2}}(-\frac{256(7+4\sqrt{2}i)}{243}x_1-\frac{1024(2+5\sqrt{2}i)}{81}x_2+\cdots)
\nonumber \\
f_4 &=&
f_1\log{x_1}+x_2^{-\frac{3}{2}}(\frac{512(23-10\sqrt{2}i)}{81}x_2+\cdots)
\end{eqnarray}

\end{enumerate}

\subsection{Analytic Continuation}

The periods of a Calabi-Yau manifold are integrals of the
holomorphic three-form over the three-cycles. In the case of the
Dijkgraaf-Vafa model, the integrals of the holomorphic three-form
over the symplectic three-cycles reduce to integrals of a
differential one-form over its branch cuts on the complex plane.
For convenience we consider the cubic potential
$W(x)=\frac{1}{2}x^2+\frac{1}{3}x^3$ with the cubic coupling $g$
set to one. The A-cycle periods and B-cycle periods are
\footnote{The periods are determined only up to a sign ambiguity
due to the square root factors in the formula. Here we have taken
the proper signs to match the convention of \cite{CIV}.}
\begin{eqnarray} \label{cycle1}
S_1&=&\frac{1}{2\pi i}\int_{x_3}^{x_4}dx
\sqrt{(x-x_1)(x-x_2)(x-x_3)(x-x_4)} \nonumber \\
S_2&=& -\frac{1}{2\pi i}\int_{x_1}^{x_2}dx
\sqrt{(x-x_1)(x-x_2)(x-x_3)(x-x_4)} \nonumber
\\ \Pi_1&=&\frac{1}{2\pi i}\int_{x_3}^{\Lambda_0}dx
\sqrt{(x-x_1)(x-x_2)(x-x_3)(x-x_4)} \nonumber
\\ \Pi_2&=&\frac{1}{2\pi i}\int_{x_2}^{\Lambda_0}dx
\sqrt{(x-x_1)(x-x_2)(x-x_3)(x-x_4)}
\end{eqnarray}

The asymptotic expansion of the periods around the origin
$(z_1,z_2)=(0,0)$ was considered in \cite{CIV}. Around this point
the roots satisfy $x_1<x_2<x_3<x_4$ and the cuts between $x_1,
x_2$ and between $x_3, x_4$ shrink to zero sizes. It was found
there that the asymptotic expansions of the periods are
\begin{eqnarray}
S_1&=&\frac{z_1}{4}-\frac{1}{8}z_1(2z_1+3z_2)+\cdots \nonumber \\
S_2&=& -\frac{z_2}{4}+\frac{1}{8}z_2(3z_1+2z_2)+\cdots \nonumber
\\
2\pi i\Pi_1&=&
\frac{\Lambda_0^3}{3}+\frac{\Lambda_0^2}{2}+S_1\log(\frac{z_1}{4})-\frac{z_1}{4}-(S_1+S_2)\log(\Lambda_0^2)+\cdots
\nonumber \\
2\pi i\Pi_2&=&
\frac{\Lambda_0^3}{3}+\frac{\Lambda_0^2}{2}-\frac{1}{6}+S_2\log(\frac{z_2}{4})+\frac{z_2}{4}-(S_1+S_2)\log(\Lambda_0^2)+\cdots
\end{eqnarray}
The Picard-Fuchs equations we derived can determine the constant
term in the B-cycle periods $\Pi_i$ and the cut-off parameter is
fixed to be $\Lambda_0=-\frac{1}{2}$. We can find the monodromy
matrices around this point $z_i\rightarrow z_ie^{2\pi i}$
\begin{eqnarray}
M_{z_1}= \left( \begin{array}{cccc}
1 & 0 & 0 & 0 \\
0 & 1 & 0 & 0\\
1 & 0 & 1 & 0 \\
0 & 0 & 0 & 1 \end{array} \right), ~~~ M_{z_2}= \left(
\begin{array}{cccc}
1 & 0 & 0 & 0 \\
0 & 1 & 0 & 0\\
0 & 0 & 1 & 0 \\
0 & 1 & 0 & 1 \end{array} \right)
\end{eqnarray}
Since the integrals are done over symplectic cycles, the monodromy
matrices are elements of the symplectic group $Sp(4,\mathcal{Z})$
and satisfy the van Kampen relation
$M_{z_1}M_{z_2}=M_{z_2}M_{z_1}$.

Now we want to analytically continue the periods to other points
in the complex structure moduli space. The analytic continuation
will fix the symplectic basis of periods, which is not available
by solving the Picard-Fuchs equation around these points. In order
to do the analytic continuation, we must do the integrals in
(\ref{cycle1}) exactly. The A-cycle periods $S_i$ and the
difference between the two B-cycle periods $\Pi_1-\Pi_2$ can be
written in terms of complete elliptic integrals of the first,
second and third kinds, and one of the B-cycle periods involves
incomplete elliptic integrals.

We consider analytically continue the periods (\ref{cycle1}) to a
singular point $(z_1,z_2)=(0,\frac{1}{3})$ in the moduli space.
This is the closest singular point to $(z_1,z_2)=(0,0)$ in the
moduli space. We will use the local coordinate
$(\tilde{z}_1,\tilde{z}_2)$ around the intersection of the blow up
divisor $F_1: \frac{1}{3}-z_2=0$ and divisor $C_1: z_1=0$ as we
did for solving the Picard-Fuchs equation
\begin{eqnarray} \label{coordinate1}
\tilde{z}_1=\frac{z_1}{(z_2-\frac{1}{3})^2}~~~
\tilde{z}_2=\frac{1}{3}-z_2
\end{eqnarray}
We directly compute the asymptotic expansion of one B-cycle period
$\Pi_2$ and use the asymptotic expansion formulae of the complete
elliptic integrals in \cite{book1} to obtain the asymptotic
formulae for other periods.

For convenience we define a function in terms of complete elliptic
integrals of the first kind $K(k^2)$, the second kind $E(k^2)$ and
the third kind $\Pi(a^2,k^2)$ as the following
\begin{eqnarray}
&&P(a^2,k^2) = \frac{1}{48 a^4 (1-a^2)^2 (a^2-k^2)^2} \nonumber \\
&\times & \{a^2[3a^8+3k^4-4a^6(1+k^2)-4a^2k^2(1+k^2)\nonumber \\
&&+ 2a^4(2+k^2+2k^4)]E(k^2)+ [-3a^{10}-3k^6+a^8(4+5k^2) \nonumber
\\ && +a^4k^2(4-10k^2)
 -2a^6(2+k^4)+3a^2k^4(1+2k^2)]K(k^2) \nonumber \\
&&+ 3[-a^{12}-5a^8k^2+5a^4k^4+k^6+2a^{10}(1+k^2)\nonumber \\
&&-2a^2k^4(1+k^2)]\Pi(a^2,k^2)\}
\end{eqnarray}

We start from the original matrix model point $(z_1,z_2)=(0,0)$ in
the complex structure moduli space where $x_1<x_2<x_3<x_4$. The
expressions for the A-cycle periods can be found using formulae in
\cite{book1}. After some algebra we found
\begin{eqnarray} \label{periods1}
S_1 &=& \frac{1}{2\pi}\int_{x_3}^{x_4} dx
\sqrt{(x_4-x)(x-x_3)(x-x_2)(x-x_1)}
\nonumber \\
&=&
\frac{1}{2\pi}\frac{2(x_4-x_3)(x_2-x_4)(x_4-x_1)^2}{\sqrt{(x_4-x_2)(x_3-x_1)}}
P(\frac{x_3-x_4}{x_3-x_1},\frac{(x_4-x_3)(x_2-x_1)}{(x_4-x_2)(x_3-x_1)})
\nonumber \\
S_2 &=& \frac{1}{2\pi}\int_{x_1}^{x_2} dx
\sqrt{(x_4-x)(x_3-x)(x_2-x)(x-x_1)}
\nonumber \\
&=&
\frac{1}{2\pi}\frac{2(x_4-x_2)(x_2-x_1)(x_3-x_2)^2}{\sqrt{(x_4-x_2)(x_3-x_1)}}
P(\frac{x_2-x_1}{x_3-x_1},\frac{(x_4-x_3)(x_2-x_1)}{(x_4-x_2)(x_3-x_1)})
\nonumber \\
\end{eqnarray}
and the difference between the two B-cycle periods is
\begin{eqnarray} \label{period2}
2\pi i(\Pi_2-\Pi_1)&=&\int_{x_2}^{x_3} dx
\sqrt{(x_4-x)(x_3-x)(x-x_2)(x-x_1)} \nonumber \\
&=&
\frac{2(x_4-x_2)(x_3-x_2)(x_2-x_1)^2}{\sqrt{(x_4-x_2)(x_3-x_1)}}
P(\frac{x_3-x_2}{x_3-x_1},\frac{(x_3-x_2)(x_4-x_1)}{(x_4-x_2)(x_3-x_1)})
\nonumber \\
\end{eqnarray}
We will take these exact formulae at the matrix model point and
analytically continue to the local coordinate (\ref{coordinate1}).

We can also directly compute the asymptotic expansion one of
B-cycle periods $\Pi_2$ around $(\tilde{z}_1,\tilde{z}_2)=(0,0)$
as follows
\begin{eqnarray} \label{periods3}
2\pi i\Pi_2 &=&
\int_{\sqrt{\frac{1}{3}-\tilde{z}_2}-I}^{\Lambda_0+\frac{1}{2}-\frac{I}{2}}
dx
\sqrt{((x+I)^2-\frac{1}{3}+\tilde{z_2})(x^2-\tilde{z}_1\tilde{z}_2^2)}
\nonumber \\ &=&
\int_{\sqrt{\frac{1}{3}-\tilde{z}_2}-I}^{\Lambda_0+\frac{1}{2}-\frac{I}{2}}
dx
\sqrt{((x+I)^2-\frac{1}{3}+\tilde{z_2})}(x-\frac{\tilde{z}_1\tilde{z}_2^2}{2x}
-\frac{(\tilde{z}_1\tilde{z}_2^2)^2}{8x^3} +O(\tilde{z}^4))
\nonumber \\
\end{eqnarray}
where
$I=\sqrt{\frac{1}{3}-2\tilde{z}_1\tilde{z}_2^2+2\tilde{z}_2}$. We
can compute the integrals exactly for the first few leading terms
written above, and expand around the cut off
$\Lambda_0=\infty$ keeping only positive powers of $\Lambda_0$.

We can now use the expressions for the periods (\ref{periods1}),
(\ref{period2}), (\ref{periods3}) and obtain the asymptotic
expansions to a first few orders
\begin{eqnarray}
S_1 &=& \frac{\sqrt{3}}{4}f_1 \nonumber \\S_2 &=&
-\frac{\sqrt{3}}{4}f_1+\frac{3\sqrt{3}}{8}f_2-\frac{1}{12\sqrt{3}}f_3
\nonumber \\
\Pi_1 &=& \Pi_2+\frac{1}{2\pi
i}\frac{\sqrt{3}}{4}(f_4+(3-\log(2^43^3))f_1)
\nonumber \\
2\pi i\Pi_2 &=&
\frac{\Lambda_0^3}{3}+\frac{\Lambda_0^2}{2}-\frac{1}{12}+
\frac{\sqrt{3}}{108}f_3-(S_1+S_2)\log(12\Lambda_0^2)
\end{eqnarray}
where $f_1, f_2, f_3, f_4$ are the asymptotic expansion of the
solutions for Picard-Fuchs equation we found earlier
\begin{eqnarray}
f_1 &=&
\tilde{z}_1\tilde{z}_2^{\frac{5}{2}}(1-\frac{1}{216}\tilde{z}_1-\frac{23}{72}\tilde{z}_1\tilde{z}_2-\frac{5}{46656}\tilde{z}_2^2+\cdots) \nonumber \\
f_2&=& \tilde{z}_2^2(1+\frac{4}{9}\tilde{z}_1+4\tilde{z}_2+\cdots) \nonumber \\
f_3 &=&
1+81\tilde{z}_2^3-108\tilde{z}_1\tilde{z}_2^3-\frac{729}{8}\tilde{z}_2^4+\cdots
\nonumber
\\
f_4&=&
f_1\log(\tilde{z}_1)+\tilde{z}_2^{\frac{5}{2}}(\frac{36}{5}-\frac{108}{7}\tilde{z}_2+\cdots)
\end{eqnarray}

These asymptotic expressions of periods are linear combinations of
the $4$ solutions to the Picard-Fuchs equations we found earlier,
provided we choose the cut-off constant to be
$\Lambda_0=-\frac{1}{2}$. Thus we have found the canonical basis
for the symplectic cycles. It is easy to write down the monodromy
matrices around this point
\begin{eqnarray} \label{monodromyF1}
M_{\tilde{z}_1}= \left( \begin{array}{cccc}
1 & 0 & 0 & 0 \\
0 & 1 & 0 & 0\\
1 & 0 & 1 & 0 \\
0 & 0 & 0 & 1 \end{array} \right), ~~~ M_{F_1}= \left(
\begin{array}{cccc}
-1 & 0 & 0 & 0 \\
2 & 1 & 0 & 0\\
0 & 0 & -1 & 2 \\
0 & 0 & 0 & 1 \end{array} \right)
\end{eqnarray}
The monodromy around the singular divisor $C_1$ is the same as
before $M_{\tilde{z}_1}=M_{z_1}$. We can also see that the
monodromy matrices are elements of $Sp(4,\mathcal{Z})$ group and
satisfy the van Kampen relation
$M_{\tilde{z}_1}M_{F_1}=M_{F_1}M_{\tilde{z}_1}$.

In general it is not easy to do the analytic continuation of
periods. We use a numerical method to match the basis of solutions
of Picard-Fuchs equation at different points of the moduli space.
We consider the intersection of the singular divisor $J$ and the
blow up divisor $F_1: \frac{1}{3}-z_2=0$. The local coordinate and
the solutions for the Picard-Fuchs equation are
\begin{eqnarray} \tilde{z}_1
=\frac{z_1(6-14z_2-9z_1)}{9(z_2-\frac{1}{3})^2}-1,~~~\tilde{z}_2=\frac{1}{3}-z_2
\end{eqnarray}

\begin{eqnarray}
f_1 &=&
\tilde{z}_1^2\tilde{z}_2^{\frac{5}{2}}(1-\frac{5}{16}\tilde{z}_1+\frac{33}{8}\tilde{z}_2+\cdots)
\nonumber \\
f_2 &=&
\tilde{z}_2^2(1+\frac{3}{4}\tilde{z}_1-\frac{43}{2}\tilde{z}_2+\cdots)
\nonumber
\\
f_3 &=& 1-648\tilde{z}_2^3+\cdots \nonumber \\
f_4 &=&
f_1\log(\tilde{z}_1)+\tilde{z}_2^{\frac{5}{2}}(\frac{512}{15}+32\tilde{z}_1-\frac{3008}{7}\tilde{z}_2+\cdots)
\end{eqnarray}

Using numerical method we find the canonical basis of the periods
as the following
\begin{eqnarray}
S_1&=&(-0.14-0.26i)f_1+0.082f_4 \nonumber \\
S_2 &=& (0.14+0.26i)f_1+2.6f_2-0.048 f_3-0.082 f_4 \nonumber \\
\Pi_1 &=&-0.26i f_1+1.03i f_2-0.016 f_3 \nonumber \\
\Pi_2 &=&1.03i f_2-0.016if_3
\end{eqnarray}
The monodromy matrix of the divisor $F_1$ is the same as we have
derived in (\ref{monodromyF1}). We can now write down the
monodromy matrix of the singular divisor $J$ by looking at the
transformation around $\tilde{z}_1\rightarrow \tilde{z}_1e^{2\pi
i}$
\begin{eqnarray}
M_{J}= \left( \begin{array}{cccc}
1 & 0 & -2 & 2 \\
0 & 1 & 2 & -2\\
0 & 0 & 1 & 0 \\
0 & 0 & 0 & 1 \end{array} \right),
\end{eqnarray}

For the singular divisor $I: 1-2z_1-2z_2=0$, we find essential
singularities instead of simple singularities. This can be seen
from the asymptotic behavior of the solutions of the Picard-Fuchs
equation at any point in the divisor. We find that the radius of
convergence for the asymptotic expansion is zero, i.e. the series
is always divergent. This is an interesting new feature of the
Dijkgraaf-Vafa model.

\section{Matrix model calculations} \label{matrixmodel}
In this Appendix we give some details of the matrix model
calculations following the approach in \cite{KMT, AKMV}. The cubic
matrix model can be expressed in the eigenvalues of the matrix
\begin{equation}
W(\Phi)=\tr(\frac{\Phi^2}{2}+\frac{g\Phi^3}{3})=\sum_{i=1}^{N}(\frac{\lambda_i^2}{2}+\frac{g\lambda_i^3}{3})
\end{equation}
Then the partition functions $Z$ and free energy $F$ are
\begin{equation}
Z=e^F=\frac{1}{Vol(U(N))}\int
D\Phi~e^{-W(\Phi)}=\frac{1}{N!(2\pi)^N}\int\prod_id\lambda_i\Delta^2(\lambda)e^{-\sum_{i=1}^{N}(\frac{\lambda_i^2}{2}+\frac{g\lambda_i^3}{3})}
\end{equation}
where $\Delta(\lambda)=\prod_{i<j}(\lambda_i-\lambda_j)$ is the
standard Verdermonde determinant from the measure of the matrix.
We expand $N_1$ eigenvalues around the critical points $a_1=0$ and
$N_2=N-N_1$ eigenvalues around the critical points
$a_2=-\frac{1}{g}$. Suppose the fluctuation is $\mu_i$, $\nu_i$
\begin{eqnarray}
\lambda_i=\mu_i,~~~~~i=1,2,\cdots N_1 \nonumber \\
\lambda_{i+N_1}=-\frac{1}{g}+\nu_i~~~~~i=1,2,\cdots N_2
\end{eqnarray}
Then the potential and the Vandermonde determinant become
\begin{equation}
W(\Phi)=\sum_{i=1}^{N_1}(\frac{\mu_i}{2}+\frac{g\mu^3}{3})-\sum_{i=1}^{N_2}(\frac{\nu_i^2}{2}-\frac{g\nu_i^3}{3})+N_2W(-\frac{1}{g})
\end{equation}
\begin{equation}
\Delta^2(\lambda)=\prod_{1\leq i_1< i_2\leq
N_1}(\mu_{i_1}-\mu_{i_2})^2 \prod_{1\leq j_1<j_2\leq
N_2}(\nu_{j_1}-\nu_{j_2})^2\prod_{1\leq i\leq N_1}\prod_{1\leq j
\leq N_2}(\mu_i-\nu_j+\frac{1}{g})^2
\end{equation}
Now we can treat the expansion around this vacuum as a model with
two matrices $\Phi_1$ with eigenvalues $\mu_i$ and $\Phi_2$ with
eigenvalues $\nu_i$. The interaction terms $\prod_{1\leq i\leq
N_1}\prod_{1\leq j \leq N_2}(\mu_i-\nu_j+\frac{1}{g})^2$ in the
Vandermonde determinant can be exponentiated and written as
potential for the two matrices, then the partition functions can
be straightforwardly evaluated by expanding the potential and
computing the expectations values of Gaussian matrix model
\cite{AKMV}. We note the fluctuation around unstable critical
point $-\frac{1}{g}$ has a wrong sign kinetic term
$-\frac{\nu_i^2}{2}$. However this model is perturbatively well
defined if we treat $\Phi_1$ as a Hermitian matrix and
analytically continue $\Phi_2$ to be a anti-Hermitian matrix.
Alternatively, one can also determine the perturbative part of the
free energy by directly evaluating the Gaussian integral for
various values of $N_1$ and $N_2$ and solving for the coefficients
in the perturbation series. Using this method we are able to push
the computations of the free energy to the eighth order, and
provide many checks of the topological string calculations in
(\ref{F2results}). The perturbative part of the free energy is the
followings
{\footnotesize
\begin{eqnarray}
&& \!\!\!\!\!\!\!\!\!\!\!\!\!\! F_{pert}= -N_2W(a_2)-2N_1N_2\log(g) \nonumber \\ &+& \Big[
(\frac{2}{3}{{N_1}}^3 - 5{{N_1}}^2{N_2} + 5{N_1}{{N_2}}^2 -
     \frac{2}{3}{{N_2}}^3)+( \frac{{N_1}}{6} - \frac{{N_2}}{6} ) \Big] g^2 \nonumber \\ &+&
      \Big[ (\frac{8}{3}{{N_1}}^4 - \frac{91}{3}{{N_1}}^3{N_2}  + 59{{N_1}}^2{{N_2}}^2 -
     \frac{91}{3}{N_1}{{N_2}}^3 + \frac{8}{3}{{N_2}}^4)+(\frac{7}{3}{{N_1}}^2 -
     \frac{31}{3}{N_1}{N_2} + \frac{7}{3}{{N_2}}^2
) \Big] g^4 \nonumber \\ &+&
  \Big[  (\frac{56}{3}{{N_1}}^5  - \frac{871}{3}{{N_1}}^4{N_2}+
     \frac{2636}{3}{{N_1}}^3{{N_2}}^2- \frac{2636}{3}{{N_1}}^2{{N_2}}^3 +
     \frac{871}{3}{N_1}{{N_2}}^4 - \frac{56}{3}{{N_2}}^5)  \nonumber \\ &&
     +( \frac{332}{9}{{N_1}}^3 -
     \frac{923}{3}{{N_1}}^2{N_2} + \frac{923}{3}{N_1}{{N_2}}^2 - \frac{332}{9}{{N_2}}^3  )
     +(\frac{35}{6}{N_1}- \frac{35}{6}{N_2}) \Big] g^6
\nonumber \\ &+&
  \Big[ (\frac{512}{3}{{N_1}}^6  - \frac{6823}{2}{{N_1}}^5{N_2}   +
     \frac{28765}{2}{{N_1}}^4{{N_2}}^2 - \frac{67310}{3}{{N_1}}^3{{N_2}}^3
     \pm \cdots) \nonumber \\ &&
     +( \frac{1864}{3}{{N_1}}^4 -
     \frac{47083}{6}{{N_1}}^3{N_2}+ 15349{{N_1}}^2{{N_2}}^2 \mp\cdots )
     +(338{{N_1}}^2- 1632{N_1}{N_2}+ 338{{N_2}}^2 ) \Big] g^8
\nonumber \\
     &+& \Big[ (  \frac{9152}{5}{{N_1}}^7- 45118{{N_1}}^6{N_2} +
     247980{{N_1}}^5{{N_2}}^2   -
     540378{{N_1}}^4{{N_2}}^3 \pm\cdots )
\nonumber \\ && +( \frac{54416}{5}{{N_1}}^5  -
     187528{{N_1}}^4{N_2} + 570066{{N_1}}^3{{N_2}}^2\mp\cdots ) \nonumber \\ &&
+(\frac{66132}{5}{{N_1}}^3 - 120880{{N_1}}^2{N_2} \pm\cdots
)+(\frac{5005}{3}{N_1}-\frac{5005}{3}{N_2} )\Big] g^{10}
\nonumber \\
    &+&
  \Big[ (
     \frac{65536}{3}{{N_1}}^8  - \frac{1933906}{3}{{N_1}}^7{N_2}   +
     \frac{13258178}{3}{{N_1}}^6{{N_2}}^2 - \frac{37761034}{3}{{N_1}}^5{{N_2}}^3  +
     \frac{52780010}{3}{{N_1}}^4{{N_2}}^4\mp\cdots)
\nonumber \\ && (\frac{1762048}{9}{{N_1}}^6 -
     \frac{12980560}{3}{{N_1}}^5{N_2} +
\frac{54863776}{3}{{N_1}}^4{{N_2}}^2 -
     \frac{256344964}{9}{{N_1}}^3{{N_2}}^3\pm\cdots)
\nonumber \\ && (\frac{1305280}{3}{{N_1}}^4-
\frac{18059582}{3}{{N_1}}^3{N_2} +
     11824166{{N_1}}^2{{N_2}}^2\mp\cdots)
\nonumber \\ && (\frac{1680704}{9}{{N_1}}^2 -
\frac{8748896}{9}{N_1}{N_2} + \frac{1680704}{9}{{N_2}}^2)
     \Big]g^{12}
\nonumber \\
&+& \Big[ (   \frac{5912192}{21}{{N_1}}^9   -
     \frac{68087967}{7}{{N_1}}^8{N_2}  + \frac{564130824}{7}{{N_1}}^7{{N_2}}^2
       - 286953520{{N_1}}^6{{N_2}}^3  + 524636640{{N_1}}^5{{N_2}}^4
       \mp\cdots) \nonumber \\ && (\frac{25136768}{7}{{N_1}}^7-
97692942{{N_1}}^6{N_2} +
     537372540{{N_1}}^5{{N_2}}^2- 1166263112{{N_1}}^4{{N_2}}^3\pm\cdots)
\nonumber \\ && (12963696{{N_1}}^5 - 244438427{{N_1}}^4{N_2} +
745362156{{N_1}}^3{{N_2}}^2 \mp\cdots ) \nonumber \\ &&
(\frac{86388296}{7}{{N_1}}^3 -
     \frac{855302550}{7}{{N_1}}^2{N_2} \pm\cdots)+(\frac{8083075}{6}{N_1} -
\frac{8083075}{6}{N_2})
     \Big] g^{14} \nonumber
\end{eqnarray}

\begin{eqnarray}
  &+&
\Big[ (  \frac{11534336}{3}{{N_1}}^{10}   -
\frac{1834216417}{12}{{N_1}}^9{N_2}  +
\frac{5978643549}{4}{{N_1}}^8{{N_2}}^2 -
     6444922816{{N_1}}^7{{N_2}}^3
\nonumber \\ && + 14743157646{{N_1}}^6{{N_2}}^4   -
19289163957{{N_1}}^5{{N_2}}^5 \pm\cdots) \nonumber \\ && (
66841600{{N_1}}^8 -
     \frac{4347551555}{2}{{N_1}}^7{N_2} +
     \frac{29785674795}{2}{{N_1}}^6{{N_2}}^2  \nonumber \\ &&-
42151343305{{N_1}}^5{{N_2}}^3 +
     58765399140{{N_1}}^4{{N_2}}^4 \mp\cdots )
\nonumber \\ && (362264064{{N_1}}^6-
\frac{35002369227}{4}{{N_1}}^5{N_2} +
\frac{148144671957}{4}{{N_1}}^4{{N_2}}^2 -
57597548553{{N_1}}^3{{N_2}}^3\pm\cdots)
\nonumber \\ && (
\frac{1882324352}{3}{{N_1}}^4  -
     \frac{28221164683}{3}{{N_1}}^3{N_2}+
     18539047948{{N_1}}^2{{N_2}}^2\mp\cdots)
\nonumber \\ && (\frac{693764720}{3}{{N_1}}^2 -
\frac{3860943680}{3}{N_1}{N_2}+\frac{693764720}{3}{{N_2}}^2)
     \Big] g^{16}
\label{eq:fmatrix}
\end{eqnarray}}
This model also contains a non-perturbative part of free energy
defined as the volume factor of the $U(N)$ gauge group as in
\cite{OV1}, where it was computed with the following result
\begin{eqnarray}
F_{n.p.} &=&
\frac{N_1^2}{2}\log(N_1)+\frac{N_2^2}{2}\log(N_2)-\frac{3}{4}(N_1^2+N_2^2)-\frac{1}{12}\log(N_1N_2)
\nonumber \\ &&
+2\zeta^{\prime}(-1)+\sum_{g=2}^{\infty}\frac{B_{2g}}{4g(g-1)}(\frac{1}{N_1^{2g-2}}+\frac{1}{N_2^{2g-2}})
\end{eqnarray}
This non-perturbative part of the matrix model has the correct
universal leading behavior of Calabi-Yau near the conifold point
of its moduli space, as first pointed out in \cite{GV} in the
context of $c=1$ string compactified at self-dual radius.


\begin{thebibliography}{10}
\newcommand{\wwwspires}{http://www.slac.stanford.edu/spires/find/hep/www}

%\cite{Bershadsky:1993cx}
\bibitem{BCOV}
  M.~Bershadsky, S.~Cecotti, H.~Ooguri and C.~Vafa,
  ``Kodaira-Spencer theory of gravity and exact results for quantum string
  amplitudes,''
  Commun.\ Math.\ Phys.\  {\bf 165}, 311 (1994)
  [arXiv:hep-th/9309140].
  %%CITATION = HEP-TH 9309140;%%

\bibitem{wqbi} E.~Witten, ``Quantum background independence in string theory,''
hep-th/9306122 .

%\cite{Dijkgraaf:2002vw}
\bibitem{DV2}
  R.~Dijkgraaf and C.~Vafa,
  ``On geometry and matrix models,''
  Nucl.\ Phys.\ B {\bf 644}, 21 (2002)
  [arXiv:hep-th/0207106].
  %%CITATION = HEP-TH 0207106;%%

%\cite{Cachazo:2001jy}
\bibitem{CIV}
  F.~Cachazo, K.~A.~Intriligator and C.~Vafa,
  ``A large N duality via a geometric transition,''
  Nucl.\ Phys.\ B {\bf 603}, 3 (2001)
  [arXiv:hep-th/0103067].
  %%CITATION = HEP-TH 0103067;%%

%\cite{Dijkgraaf:2002fc}
\bibitem{DV1}
  R.~Dijkgraaf and C.~Vafa,
  ``Matrix models, topological strings, and supersymmetric gauge theories,''
  Nucl.\ Phys.\ B {\bf 644}, 3 (2002)
  [arXiv:hep-th/0206255].
  %%CITATION = HEP-TH 0206255;%%

\bibitem{GV}
  R.~Gopakumar and C.~Vafa,
  ``On the gauge theory/geometry correspondence,''
  Adv.\ Theor.\ Math.\ Phys.\  {\bf 3} (1999) 1415
  [arXiv:hep-th/9811131].

%\cite{Aganagic:2002wv}
\bibitem{AKMV}
  M.~Aganagic, A.~Klemm, M.~Marino and C.~Vafa,
  ``Matrix model as a mirror of Chern-Simons theory,''
  JHEP {\bf 0402}, 010 (2004)
  [arXiv:hep-th/0211098].
  %%CITATION = HEP-TH 0211098;%%

%\cite{Klemm:2002pa}
\bibitem{KMT}
  A.~Klemm, M.~Marino and S.~Theisen,
  ``Gravitational corrections in supersymmetric gauge theory and matrix
  models,''
  JHEP {\bf 0303}, 051 (2003)
  [arXiv:hep-th/0211216].
  %%CITATION = HEP-TH 0211216;%%

\bibitem{Aganagic:2003db}
  M.~Aganagic, A.~Klemm, M.~Marino and C.~Vafa,
  %``The topological vertex,''
  Commun.\ Math.\ Phys.\  {\bf 254} (2005) 425
  [arXiv:hep-th/0305132].
  %%CITATION = HEP-TH 0305132;%%. However


%\cite{Dijkgraaf:2002yn}
\bibitem{DST}
  R.~Dijkgraaf, A.~Sinkovics and M.~Temurhan,
  ``Matrix models and gravitational corrections,''
  Adv.\ Theor.\ Math.\ Phys.\  {\bf 7}, 1155 (2004)
  [arXiv:hep-th/0211241].
  %%CITATION = HEP-TH 0211241;%%


%\cite{Vasiliev:2005qj}
\bibitem{Vasiliev:2005qj}
  D.~Vasiliev,
  ``Determinant formulas for matrix model free energy,''
  arXiv:hep-th/0506155.
  %%CITATION = HEP-TH 0506155;%%

%\cite{Ambjorn:1992gw}
\bibitem{ACKM}
  J.~Ambjorn, L.~Chekhov, C.~F.~Kristjansen and Y.~Makeenko,
  ``Matrix model calculations beyond the spherical limit,''
  Nucl.\ Phys.\ B {\bf 404}, 127 (1993)
  [Erratum-ibid.\ B {\bf 449}, 681 (1995)]
  [arXiv:hep-th/9302014].
  %%CITATION = HEP-TH 9302014;%%



%\cite{Akemann:1996zr}
\bibitem{Akemann}
  G.~Akemann,
  ``Higher genus correlators for the Hermitian matrix model with multiple
  cuts,''
  Nucl.\ Phys.\ B {\bf 482}, 403 (1996)
  [arXiv:hep-th/9606004].
  %%CITATION = HEP-TH 9606004;%%

\bibitem{Moore:1997pc}
  G.~W.~Moore and E.~Witten,
  ``Integration over the u-plane in Donaldson theory,''
  Adv.\ Theor.\ Math.\ Phys.\  {\bf 1}, 298 (1998)
  [arXiv:hep-th/9709193].
  %%CITATION = HEP-TH 9709193;%%



\bibitem{Nekrasov:2002qd}
  N.~A.~Nekrasov,
  ``Seiberg-Witten prepotential from instanton counting,''
  Adv.\ Theor.\ Math.\ Phys.\  {\bf 7}, 831 (2004)
  [arXiv:hep-th/0206161].
  %%CITATION = HEP-TH 0206161;%%.

\bibitem{Flume:2002az}
  R.~Flume and R.~Poghossian,
  ``An algorithm for the microscopic evaluation of the coefficients of the
  Seiberg-Witten prepotential,''
  Int.\ J.\ Mod.\ Phys.\ A {\bf 18}, 2541 (2003)
  [arXiv:hep-th/0208176].
  %%CITATION = HEP-TH 0208176;%%

\bibitem{Nakajima2003} H.~Nakajima, K.~Yoshioka,
    ``Instanton counting on blowup. I.
     4-dimensional pure gauge theory,''
      math.AG/0306198.

\bibitem{Nekrasov:2003rj}
  N.~Nekrasov and A.~Okounkov,
  ``Seiberg-Witten theory and random partitions,''
  arXiv:hep-th/0306238.

\bibitem{Seiberg:1994rs}
  N.~Seiberg and E.~Witten,
  ``Electric - magnetic duality, monopole condensation, and confinement in N=2
  supersymmetric Yang-Mills theory,''
  Nucl.\ Phys.\ B {\bf 426}, 19 (1994)
  [Erratum-ibid.\ B {\bf 430}, 485 (1994)]
  [arXiv:hep-th/9407087].
  %%CITATION = HEP-TH 9407087;%%

%\cite{Katz:1996fh}
\bibitem{KKV}
  S.~Katz, A.~Klemm and C.~Vafa,
  ``Geometric engineering of quantum field theories,''
  Nucl.\ Phys.\ B {\bf 497}, 173 (1997)
  [arXiv:hep-th/9609239].
  %%CITATION = HEP-TH 9609239;%%

\bibitem{Klemm:1995wp}
  A.~Klemm, W.~Lerche and S.~Theisen,
  ``Nonperturbative effective actions of N=2 supersymmetric gauge theories,''
  Int.\ J.\ Mod.\ Phys.\ A {\bf 11} (1996) 1929
  [arXiv:hep-th/9505150].
  %%CITATION = HEP-TH 9505150;%%

\bibitem{Nahm:1996di}
  W.~Nahm,
  ``On the Seiberg-Witten approach to electric-magnetic duality,''
  arXiv:hep-th/9608121.
  %%CITATION = HEP-TH 9608121;%%

\bibitem{Klein} F.~Klein, {\sl Vorlesung \"uber die Theorie der elliptischen
Modulfunktionen,} Teubner Leipzig (1890)\ .

\bibitem{Ferrara:2006js}
  S.~Ferrara and O.~Macia,
  %``Observations on the Darboux coordinates for rigid special geometry,''
  arXiv:hep-th/0602262.
  %%CITATION = HEP-TH 0602262;%%

\bibitem{Bershadsky:1993ta}
  M.~Bershadsky, S.~Cecotti, H.~Ooguri and C.~Vafa,
  ``Holomorphic anomalies in topological field theories,''
  Nucl.\ Phys.\ B {\bf 405}, 279 (1993)
  [arXiv:hep-th/9302103].
  %%CITATION = HEP-TH 9302103;%%

\bibitem{Klemm:1999gm}
  A.~Klemm and E.~Zaslow,
  ``Local mirror symmetry at higher genus,''
  arXiv:hep-th/9906046.
  %%CITATION = HEP-TH 9906046;%%


\bibitem{marinoperiods}
  L.~Alvarez-Gaume, M.~Marino and F.~Zamora,
  ``Softly broken N = 2 {QCD} with massive quark hypermultiplets. I,''
  Int.\ J.\ Mod.\ Phys.\ A {\bf 13}, 403 (1998)
  [arXiv:hep-th/9703072].
  %%CITATION = HEP-TH 9703072;%%

\bibitem{marcosreview} M.\ Marino,
``The uses of Whitham hierachies'', 
  arXiv:hep-th/9905053.

\bibitem{Matone} M.~Matone,
  ``Instantons and recursion relations in N=2 SUSY gauge theory,'' 
   Phys.\ Lett.\ B {\bf 357}, 342 (1995)[arXiv:hep-th/9506102] and 
  ``Koebe 1/4 theorem and inequalities in N=2 superQCD,''
   Phys.\ Rev.\ D {\bf 53}, 7354 (1996)
  [arXiv:hep-th/9506181].

\bibitem{Witten:1991zd}
  E.~Witten,
  %``Ground ring of two-dimensional string theory,''
  Nucl.\ Phys.\ B {\bf 373}, 187 (1992)
  [arXiv:hep-th/9108004].
  %%CITATION = HEP-TH 9108004;%%

\bibitem{DVIII}
  R.~Dijkgraaf and C.~Vafa,
  ``A perturbative window into non-perturbative physics,''
  arXiv:hep-th/0208048.
  %%CITATION = HEP-TH 0208048;%%

\bibitem{Dijkgraaf} R.~Dijkgraaf, ``Mirror symmetry and elliptic curves,''
 in {\sl The Moduli Space of Curves,} Progr. Math. {\bf 129}
    (Birkh\"auser, 1995), 149.

\bibitem{Zagier} K.~Kaneko and D.~B.~Zagier, ``A generalized Jacobi
theta function and quasi-modular form,'' ididem, 165.

\bibitem{Borcherds}
        R.~E.~Borcherds,
        ``Automorphic forms with singularities on Grassmannians,''
        Invent. Math. {\bf 132} (1998) 491--562
        [arXiv:alg-geom/9609022].

\bibitem{kontsevich1}
        M.~Kontsevich,
        ``Product formulas for modular forms on O(2,n) (after R.Borcherds),''
        Ast\'erisque {\bf 245} (1997), Exp. No. 821, 41--56.
        [arXiv:alg-geom/9709006].

\bibitem{Kawai}
        T.~Kawai and K.~Yoshioka,
        ``String Partition function and Infinite Products,''
        [arXiv:hep-th/0002169].

\bibitem{hosono}
        S.~Hosono, M.~H.~Saito, and A.~Takahashi,
        ``Holomorphic anomaly equation and BPS state counting of
        rational  elliptic surface,''
        Adv.\ Theor.\ Math.\ Phys.\ {\bf 3} (1999) 177 -- 208
        [arXiv:hep-th/9901151].

\bibitem{KKRS} A.~Klemm, M.~Kreuzer, E.~Riegler and E.~Scheidegger,
          ``Topological string amplitudes, complete intersection Calabi-Yau spaces and
          threshold corrections,''
          JHEP {\bf 0505}, 023 (2005)
          [arXiv:hep-th/0410018].
          %%CITATION = HEP-TH 0410018;%%


\bibitem{Klemm:2005pd}
  A.~Klemm and M.~Marino,
  ``Counting BPS states on the Enriques Calabi-Yau,''
  arXiv:hep-th/0512227.
  %%CITATION = HEP-TH 0512227;%%

\bibitem{Ghoshal:1995wm}
  D.~Ghoshal and C.~Vafa,
  ``C = 1 string as the topological theory of the conifold,''
  Nucl.\ Phys.\ B {\bf 453}, 121 (1995)
  [arXiv:hep-th/9506122].
  %%CITATION = HEP-TH 9506122;%%

%\cite{Ooguri:2003qp}
\bibitem{OV2}
  H.~Ooguri and C.~Vafa,
  ``The C-deformation of gluino and non-planar diagrams,''
  Adv.\ Theor.\ Math.\ Phys.\  {\bf 7}, 53 (2003)
  [arXiv:hep-th/0302109].
  %%CITATION = HEP-TH 0302109;%%

%\cite{Ooguri:2003tt}
\bibitem{OV3}
  H.~Ooguri and C.~Vafa,
  ``Gravity induced C-deformation,''
  Adv.\ Theor.\ Math.\ Phys.\  {\bf 7}, 405 (2004)
  [arXiv:hep-th/0303063].
  %%CITATION = HEP-TH 0303063;%%

\bibitem{DGOVZ}
  R.~Dijkgraaf, M.~T.~Grisaru, H.~Ooguri, C.~Vafa and D.~Zanon,
  ``Planar gravitational corrections for supersymmetric gauge theories,''
  JHEP {\bf 0404}, 028 (2004)
  [arXiv:hep-th/0310061].
  %%CITATION = HEP-TH 0310061;%%

\bibitem{KM}
  A.~Klemm and M.~Marino,
  ``Counting BPS states on the Enriques Calabi-Yau,''
  arXiv:hep-th/0512227.
  %%CITATION = HEP-TH 0512227;%%

%\cite{Cecotti:1991me}
\bibitem{CV}
  S.~Cecotti and C.~Vafa,
  ``Topological antitopological fusion,''
  Nucl.\ Phys.\ B {\bf 367}, 359 (1991).
  %%CITATION = NUPHA,B367,359;%%


\bibitem{book1}
P.~F.~Byrd and M.~D.~Friedman, ``Handbook of Elliptic Integrals
for Engineers and Scientists'', Springer-Verlag, 1971.


%\cite{Ooguri:2002gx}
\bibitem{OV1}
  H.~Ooguri and C.~Vafa,
  ``Worldsheet derivation of a large N duality,''
  Nucl.\ Phys.\ B {\bf 641}, 3 (2002)
  [arXiv:hep-th/0205297].
  %%CITATION = HEP-TH 0205297;%%


\bibitem{Chekhov:2005rr}
  L.~Chekhov and B.~Eynard,
  ``Hermitean matrix model free energy: Feynman graph technique for all
  %genera,''
  JHEP {\bf 0603}, 014 (2006)
  [arXiv:hep-th/0504116].
  %%CITATION = HEP-TH 0504116;%%



\bibitem{Chekhov:2006rq}
  L.~Chekhov and B.~Eynard,
  ``Matrix eigenvalue model: Feynman graph technique for all genera,''
  arXiv:math-ph/0604014.
  %%CITATION = MATH-PH 0604014;%%

\bibitem{ABK}
M.~Aganagic, V.~Bouchard and A.~Klemm, ``Topological Strings and (Almost) Modular Forms,''
  arXiv:hep-th/0607100.

\bibitem{workinprogress} Work in progress.


\end{thebibliography}
\end{document}